\documentclass[11pt,a4paper]{article}
\usepackage{jheppub}
\usepackage{doi}
\usepackage{hyperref}
\usepackage{amsmath}
\usepackage{amssymb}
\usepackage{mathrsfs}
\usepackage{graphicx}
\usepackage{subcaption}
\usepackage{dcolumn}
\usepackage{upgreek}
\usepackage{multirow}
\usepackage{orcidlink}
\usepackage{url}
\usepackage[toc,page]{appendix}
\usepackage[normalem]{ulem}
\usepackage{pdfpages}

\usepackage{feynmp-auto}
\usepackage{color}
\usepackage[compat=1.1.0]{tikz-feynman}
\usepackage{xcolor}

\renewcommand{\arraystretch}{1.4}

\newcommand{\eq}{\begin{eqnarray}}
\newcommand{\en}{\end{eqnarray}}
\newcommand{\la}{\langle}
\newcommand{\ra}{\rangle}
\newcommand{\ctau}{{c\tau_{\tilde{\chi}^0_1}}}
\newcommand{\neu}{{\tilde{\chi}_1^0}}
\newcommand{\order}[1]{\mathcal{O}{(#1)}}
\newcommand{\DV}{\text{DV}}
\newcommand{\B}{\mathcal{B}}

\title{Searching for long-lived light neutralinos from $B$-meson decays with baryonic R-parity violation at Belle II}

\author[b]{Emilie Bertholet\,\orcidlink{0000-0002-3792-2450},}
\emailAdd{ebertholet@tauex.tau.ac.il}

\author[c]{Claudio O.~Dib\,\orcidlink{0000-0003-4146-906X},}
\emailAdd{claudio.dib@usm.cl}

\author[b]{Sara P.~Gandelman\,\orcidlink{0009-0003-4682-3213},}
\emailAdd{sarag@mail.tau.ac.il}

\author[d,e]{Juan Carlos Helo\,\orcidlink{0000-0002-5310-8598},}
\emailAdd{jchelo@userena.cl}

\author[f,e]{Valery~E.~Lyubovitskij\,\orcidlink{0000-0001-7467-572X},}
\emailAdd{valeri.lyubovitskij@uni-tuebingen.de}

\author[g]{Minakshi Nayak\,\orcidlink{0000-0002-2572-4692},}
\emailAdd{minakshin@iisc.ac.in}

\author[h]{Nicol\'as A.~Neill\,\orcidlink{0000-0001-7897-5834},}
\emailAdd{naneill@outlook.com}

\author[b]{Abner Soffer\,\orcidlink{0000-0002-0749-2146},}
\emailAdd{asoffer@tau.ac.il}

\author[a]{Zeren Simon Wang\,\orcidlink{0000-0002-1483-6314}}
\emailAdd{wzs@hfut.edu.cn}

\affiliation[a]{School of Physics, Hefei University of Technology, Hefei 230601, China}
\affiliation[b]{School of Physics and Astronomy, Tel Aviv University, Tel Aviv 69978, Israel}
\affiliation[c]{Departmento de F\'isica and CCTVal, Universidad T\'ecnica Federico Santa Mar\'ia,\\ Valpara\'iso 2340000, Chile}
\affiliation[d]{Departamento de F\'{i}sica, Facultad de Ciencias, Universidad de La Serena,\\
Avenida Cisternas 1200, La Serena, Chile}
\affiliation[e]{Millennium Institute for Subatomic Physics at the High Energy Frontier (SAPHIR), Fern\'andez Concha 700, Santiago, Chile}
\affiliation[f]{Institut f\"ur Theoretische Physik, Universit\"at T\"ubingen, \\
		Kepler Center for Astro and Particle Physics, \\ 
		Auf der Morgenstelle 14, D-72076 T\"ubingen, Germany}
\affiliation[g]{Centre for High Energy Physics, Indian Institute of Science, Bangalore 560012, India}
\affiliation[h]{Centro Multidisciplinario de F\'isica, Vicerrector\'ia de Investigaci\'on, Universidad Mayor, 8580745 Santiago, Chile}

\abstract{In a supersymmetry scenario with R-parity violation (RPV), neutralinos with GeV-scale mass, which are necessarily bino-like, are allowed by all constraints and can be produced in association with a baryon in $B$-meson decays via certain $\bar U \bar D \bar D$ operators.
In this work, we investigate this scenario with two non-vanishing RPV couplings at the low-energy scale.
With one RPV coupling governing the neutralino production rate and another determining its lifetime, this scenario can lead to observable signals with displaced-vertex signatures in the tracking volume of $B$-factories.
To maximize the sensitivity to such signals, we develop a new partial-reconstruction technique that yields high efficiency and utilizes most of the decays of relatively heavy, long-lived particles, achieving much better sensitivity than standard full reconstruction.
We consider potential background sources and devise selection criteria to suppress their event yields to very low levels. 
Using a parameterized model of the detector, we estimate in detail the displaced-vertex reconstruction efficiency as a function of neutralino lifetime and mass.
For squark masses beyond the LHC limits, we calculate the signal sensitivity of Belle~II, showing that the experiment can probe the RPV couplings well beyond the present bounds, obtained from searches for dinucleon decays, baryon-antibaryon oscillations, and $B^+\to p +\text{missing}$.
}

\begin{document}
\maketitle

\section{Introduction}
\label{sec:intro}

While the discovery of a Standard-Model (SM) like Higgs boson~\cite{ATLAS:2012yve,CMS:2012qbp} has completed the SM spectrum, there remain multiple theoretical and observational issues associated with the SM, indicating that it cannot be the complete picture of fundamental physics.
To date, supersymmetry (SUSY) (for reviews see Refs.~\cite{Nilles:1983ge,Martin:1997ns}) remains one of the leading theoretical ideas for physics beyond the SM; by proposing a fermion-boson symmetry, it not only solves the hierarchy problem~\cite{Gildener:1976ai,Veltman:1980mj} but also entails other desirable features, such as predicting unification of gauge couplings at the Grand-Unified-Theory scale.
SUSY predicts new heavy fields that have been searched for intensively at the LHC for more than 15 years.
While no concrete discovery of such particles has been made, stringent lower bounds on their masses have been established.
For instance, the masses of the squarks have been bounded from below at 1.85 or 1.22~TeV, depending on the assumed number of degenerate degrees of freedom~\cite{Beringer:2024ady,ATLAS:2018nud,CMS:2017brl,CMS:2019vzo,CMS:2019zmd,ATLAS:2020xgt}.
Even the color-neutral sleptons have mass bounds in the range of several hundred GeV~\cite{ATLAS:2022hbt,CMS:2023qhl}.
On the other hand, the bounds on the masses of the predicted neutralinos and charginos are much looser.
In fact, it has been shown that if one lifts the GUT relation  $m_1=\frac{5}{3}\tan^2{\theta_W}m_2$ between the gaugino masses $m_1$, $m_2$, where $\theta_W$ is the electroweak mixing angle, and the lightest neutralino $\neu$ is dominantly bino-like and obeys the current laboratory bounds~\cite{Gogoladze:2002xp,Dreiner:2009ic} and astrophysical constraints~\cite{Dreiner:2003wh,Kachelriess:2000dz,Dreiner:2013tja}, a $\neu$ mass as light as in the GeV-scale is allowed~\cite{Choudhury:1995pj,Choudhury:1999tn,Belanger:2002nr,Bottino:2002ry,Belanger:2003wb,AlbornozVasquez:2010nkq,Calibbi:2013poa,Gogoladze:2002xp,Dreiner:2009ic} as long as it decays to avoid overclosing the Universe~\cite{Hooper:2002nq,Bottino:2011xv,Belanger:2013pna,Bechtle:2015nua}.

We note that LHC searches do not constrain the scenario of a GeV-scale bino-like neutralino as the lightest supersymmetric particle (LSP) with all other sparticle masses being beyond the current LHC reach.
For example, current constraints on neutralino production from searches for $pp\to \tilde q\tilde q\to qq\neu\neu$, giving rise to jets plus missing transverse energy, are not relevant for squark masses above the current LHC limit, $m_{\tilde q}> 1.85$~TeV~\cite{atlas:2020syg}.
Moreover, the LHC measurements of the Higgs parameters do not exclude our scenario.
This is the case since, in the MSSM, the coupling of the SM-like Higgs boson $h$ to a pair of neutralinos requires the Higgsino component, while we consider the lightest neutralino to be purely bino-like.
As a result, the decay amplitude of $h\to \tilde{\chi}^0_1\tilde{\chi}^0_1$ is vanishing.
Therefore, we conclude that such a light neutralino remains largely unconstrained at present, as long as we take $m_{\tilde q}> 1.85$~TeV~\cite{atlas:2020syg}.

One theoretical possibility for the lightest neutralino to decay is via R-parity violation (RPV)~\cite{Dreiner:1997uz,Barbier:2004ez,Mohapatra:2015fua,Allanach:2003eb}.
RPV SUSY is theoretically as legitimate as its usual R-parity-conserving (RPC) counterpart, yet provides a much richer collider phenomenology (see for example Refs.~\cite{Dreiner:1991pe,deCampos:2007bn,Dercks:2017lfq,Dreiner:2023bvs,Dreiner:2025kfd}). 
The RPV SUSY includes sets of operators in the superpotential that violate either lepton number or baryon number.
Switching on both types of operators to the level accessible by collider experiments would lead to proton-decay rates in conflict with the current bounds on the proton lifetime~\cite{Beringer:2024ady,Chamoun:2020aft}. 
We avoid this problem by focusing on the case where only the baryon-number-violating (BNV) operator terms $\lambda'' \bar U \bar D \bar D$ are non-vanishing, which can be realized by, e.g., imposing the proton hexality symmetry~\cite{Dreiner:2005rd,Dreiner:2007vp}.
Although in RPV SUSY, any supersymmetric particle can be the LSP~\cite{Dercks:2017lfq,Dreiner:2008ca,Desch:2010gi}, we take the lightest neutralino $\neu$ to be the LSP in this work.
Decaying via a small RPV coupling into SM particles, a light neutralino is naturally long-lived, leading to exotic signatures of displaced vertices (DVs) in laboratory experiments for a range of  RPV coupling and neutralino mass values.

At the ongoing $B$-factory experiment Belle~II~\cite{Belle-II:2010dht,Belle-II:2018jsg}, as many as $\order{10^{10}}$ $B$-mesons are projected to be produced with the anticipated integrated luminosity of 50 ab$^{-1}$.
A $B$-meson can undergo the decay $B\to \neu {\cal B}$ to the neutralino $\neu$ and a baryon $ {\cal B}$ via a BNV coupling. 
In Ref.~\cite{Dib:2022ppx}, some of us investigated the resulting $B$-factories' sensitivity to a GeV-scale neutralino with a single non-vanishing BNV coupling, a scenario explored experimentally by the BABAR collaboration~\cite{BaBar:2023rer,BaBar:2023dtq}.
In that case, the vast majority of neutralino decays take place outside the detector, so that the neutralino signature is ``missing-energy''.
Therefore, neutralino identification requires full reconstruction of the event's other $B$-meson in a hadronic final state. 
The sensitivity of this technique is limited by the small efficiency of the full reconstruction efficiency, which is less than 1\%.

In the current work, in addition to neutralino production via $B\to \neu {\cal B}$, we study the case of a second non-vanishing BNV coupling, which facilitates the neutralino decay into a state with baryon number $\pm 1$.
For a range of values of this coupling, the neutralino decay can give rise to an exotic DV signature, which can be reconstructed with higher efficiency and less background than the ``baryon + missing-energy'' signature studied in Ref.~\cite{Dib:2022ppx}.
We devise a novel partial-reconstruction technique that exploits the DV signature to utilize most of the neutralino decay modes, resulting in high sensitivity relative to traditional full-reconstruction techniques. 
We evaluate the detection efficiencies as functions of the neutralino mass and lifetime with Monte-Carlo (MC) simulations and a parameterized model of the Belle~II detector.
Finally, we calculate the sensitivity of Belle~II to the  light-neutralino signal in this scenario.

For additional studies of a light neutralino with RPV SUSY at Belle~II, we refer the readers to Refs.~\cite{Nelson:2019fln,Alonso-Alvarez:2019fym,Dey:2020juy,Wang:2023trd}.
In particular, Refs.~\cite{Nelson:2019fln,Alonso-Alvarez:2019fym} studied model setups similar to ours that predict signatures of $B$-meson decays into a baryon and missing energy or displaced vertex, motivated by explaining either dark matter or the matter-antimatter asymmetry of the Universe.

This work is structured as follows.
In Sec.~\ref{sec:theory}, we introduce the RPV SUSY model and the signal processes, including a summary of the theoretical benchmark scenarios we study. 
We then compute the neutralino production and decay rates and provide an overview of the current bounds on the BNV couplings of interest.
In Sec.~\ref{sec:techniqueANDbackground} we introduce the Belle~II detector, explain the experimental techniques of partial and full reconstruction, elaborate on methods for background suppression and estimation, and describe the method for reconstruction-efficiency calculation.
We present and discuss the numerical results in Sec.~\ref{sec:results} and provide a summary and outlook in Sec.~\ref{sec:conclusions}.
We note that our results focus on neutralino production via $B^+\to \neu p$.
For calculations and conclusions involving other final-state baryons, the reader is referred to Ref.~\cite{Bertholet:2025lcr}.
Additionally, we provide the expressions for the relevant hadronic-transition form factors in Appendix~\ref{app:formfactors}, and the neutralino-candidate mass calculation with the partial-reconstruction method in Appendix~\ref{app:mass}.

\section{Theoretical model and signal process}
\label{sec:theory}

As discussed in the previous section, we focus on a light, GeV-scale neutralino as the LSP, in the theoretical framework of the RPV SUSY with only BNV interactions $\lambda'' \bar U \bar D \bar D$.
The relevant terms in the model Lagrangian are given here as the sum of two parts:
\begin{eqnarray}
    \mathcal{L}_{\text{int.}} &=& \mathcal{L}_{\text{RPC}} + \mathcal{L}_{\text{RPV}} \, ,  \label{eq:full_Lagrangian} \\
\mathcal{L}_{\text{RPC}} &\supset& - \sum\limits_{q=u,d,s,c,b} \, g^{\tilde q}_{1R} 
\, \left(\bar q_{R,a} P_L \neu\right) \tilde q_{R,a} + {\mathrm{h.c.}} \,,   \label{eq:RPC_Lagrangian}\\
\mathcal{L}_{\text{RPV}} &=&  \sum_{i,j,k=1}^3\lambda''_{ijk}\, \epsilon_{abc}   \left(
        \tilde u^{*}_{Ria}\, \bar d_{Rjb}\, d^{C}_{Rkc}      
      + \tilde d^{*}_{Rja}\,  \bar u_{Rib}\,  d^{C}_{Rkc}
      + \tilde d^{*}_{Rka}\,  \bar u_{Rib}\,  d^{C}_{Rjc} 
        \right) + {\mathrm{h.c.}}, \label{eq:RPV_Lagrangian}
\end{eqnarray}
where $\mathcal{L}_{\text{RPC}}$ includes the RPC interactions associating the light neutralino (taken to be a pure bino) with a quark and a squark; $\mathcal{L}_{\text{RPV}}$ lists the BNV RPV terms between a squark and two quarks; $\lambda''_{ijk}$ are dimensionless BNV RPV couplings that vanish for $j=k$; $a,b,c$ are color indices; $\epsilon_{abc}$ is the three-dimensional anti-symmetric tensor; $P_L=\frac{1-\gamma^5}{2}$ is the left-chiral projector; the superscript $C$ labels charge conjugation; and the subscript $R$ indicates a right-chiral field or coupling.
For simplicity, we neglect squark mixings, so that only right-chiral fields are considered.
Taking $\neu$ to be a pure bino, the coupling $g_{1R}^{\tilde{q}}$ is
\begin{eqnarray}
g^{\tilde q}_{1R} = - \sqrt{2} g_W \, e_q \, \tan\theta_W   \,,\label{eqn:bino_coupling}
\end{eqnarray} 
with the tangent of the electroweak mixing angle $\tan\theta_W \simeq 0.55$,  the $SU(2)$ gauge coupling $g_W = e/\sin\theta_W \simeq 0.63$, and the quark electric charge $e_q$.

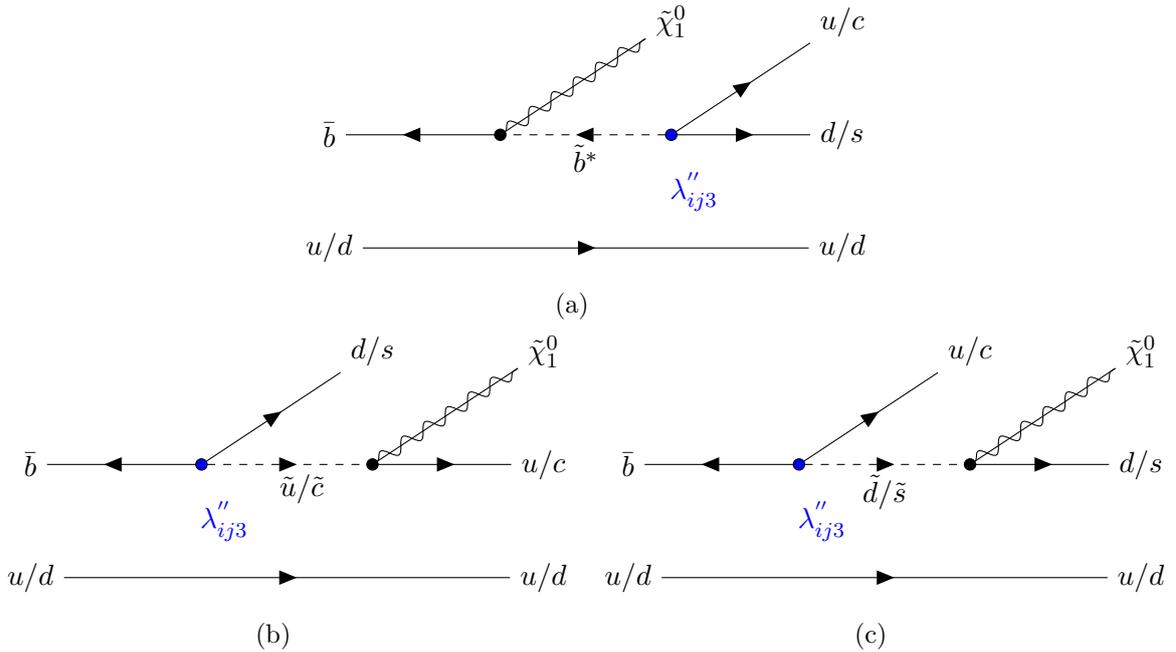
\begin{figure}[t]
    \centering

    \begin{minipage}{\textwidth}
    \centering
\begin{subfigure}[b]{0.48\textwidth}
    \centering
    \begin{tikzpicture}[scale=1.5]
        \tikzfeynmanset{every vertex={dot, minimum size=2pt}}
        \begin{feynman}
            \vertex (i1) at (0, 0) {\(\bar b\)};
            \vertex (i2) at (0, -1) {\(u/d\)};
            \vertex [dot] (v1) at (1.5, 0) {};
            \vertex [dot, fill=blue] (v2) at (3, 0) {};
            \vertex (o1) at (4.5, -1) {\(u/d\)};
            \vertex (o2) at (4.5, 0) {\(d/s\)};
            \vertex (o3) at (4.5, 1) {\(u/c\)};
            \vertex (o4) at (3, 1) {\(\tilde{\chi}_1^0\)};

            \diagram* {
                (i1) -- [anti fermion] (v1),
                (v1) -- [solid] (o4),
                (v1) -- [decorate, decoration={snake}, draw=black] (o4),
                (v2) -- [scalar, dashed, with arrow=0.5, edge label=\(\tilde b^*\)] (v1),
                (v2) -- [fermion] (o2),
                (v2) -- [fermion] (o3),
                (i2) -- [fermion] (o1),
            };

            \node[blue] at (3.2, -0.5) {\(\lambda^{''}_{ij3}\)};
        \end{feynman}
    \end{tikzpicture}
    \caption{}
\end{subfigure}
    \end{minipage}

    \begin{minipage}{\textwidth}
    \centering
    
\begin{subfigure}[b]{0.48\textwidth}
    \centering
    \begin{tikzpicture}[scale=1.5]
        \tikzfeynmanset{every vertex={dot, minimum size=2pt}}
        \begin{feynman}
            \vertex (i1) at (0, 0) {\(\bar b\)};
            \vertex (i2) at (0, -1) {\(u/d\)};
            \vertex [dot, fill=blue] (v1) at (1.5, 0) {};
            \vertex [dot] (v2) at (3, 0) {};
            \vertex (o1) at (4.5, -1) {\(u/d\)};
            \vertex (o2) at (4.5, 0) {\(u/c\)};
            \vertex (o3) at (4.5, 1) {\(\tilde{\chi}_1^0\)};
            \vertex (o4) at (3, 1) {\(d/s\)};

            \diagram* {
                (i1) -- [anti fermion] (v1),
                (v1) -- [fermion] (o4),
                (v1) -- [scalar, dashed, with arrow=0.5, edge label'=\(\quad\tilde u/\tilde c\)] (v2),
                (v2) -- [solid] (o3),
                (v2) -- [decorate, decoration={snake}, draw=black] (o3),
                (v2) -- [fermion] (o2),
                (i2) -- [fermion] (o1),
            };
            \node[blue] at (1.7, -0.5) {\(\lambda^{''}_{ij3}\)};
        \end{feynman}
    \end{tikzpicture}
    \caption{}
\end{subfigure}
    \hfill
\begin{subfigure}[b]{0.48\textwidth}
    \centering
    \begin{tikzpicture}[scale=1.5]
        \tikzfeynmanset{every vertex={dot, minimum size=2pt}}
        \begin{feynman}
            \vertex (i1) at (0, 0) {\(\bar b\)};
            \vertex (i2) at (0, -1) {\(u/d\)};
            \vertex [dot, fill=blue] (v1) at (1.5, 0) {};
            \vertex [dot] (v2) at (3, 0) {};
            \vertex (o1) at (4.5, -1) {\(u/d\)};
            \vertex (o2) at (4.5, 0) {\(d/s\)};
            \vertex (o3) at (4.5, 1) {\(\tilde{\chi}_1^0\)};
            \vertex (o4) at (3, 1) {\(u/c\)};

            \diagram* {
                (i1) -- [anti fermion] (v1),
                (v1) -- [fermion] (o4),
                (v1) -- [scalar, dashed, with arrow=0.5, edge label'=\(\tilde d/\tilde s\)] (v2),
                (v2) -- [solid] (o3),
                (v2) -- [decorate, decoration={snake}, draw=black] (o3),
                (v2) -- [fermion] (o2),
                (i2) -- [fermion] (o1),
            };
            \node[blue] at (1.7, -0.5) {\(\lambda^{''}_{ij3}\)};
        \end{feynman}
    \end{tikzpicture}
    \caption{}
\end{subfigure}

    \end{minipage}
    \caption{Parton-level diagrams for the decays $B^{+/0} \rightarrow {\cal B}\tilde{\chi}_1^0$, where $\B$ represents one baryon, are shown with the corresponding RPV coupling \( \lambda''_{ij3} \).
    The possible baryons are: \( p, n \) (for \( \lambda''_{113} \)); \( \Lambda, \Sigma^+, \Sigma^0 \) (for \( \lambda''_{123} \)); \( \Lambda^+_c, \Sigma^+_c, \Sigma^0_c \) (for \( \lambda''_{213} \)); and \( \Xi^+_c, \Xi^0_c \) (for \( \lambda''_{223} \)).
    }
    \label{fig:production-diagram}
\end{figure}

The interaction terms in the Lagrangian can lead to decays of a $B$-meson into a baryon and a neutralino, shown in Fig.~\ref{fig:production-diagram}.
In this paper, we restrict the discussion to the RPV coupling $\lambda''_{113}$ and hence the decay $B^+\to p \neu$. 
We refer the reader to Ref.~\cite{Dib:2022ppx} for the calculation of this decay's width, as well as its comparison to decays that are mediated by other $\lambda''_{ij3}$ couplings and produce other baryons.

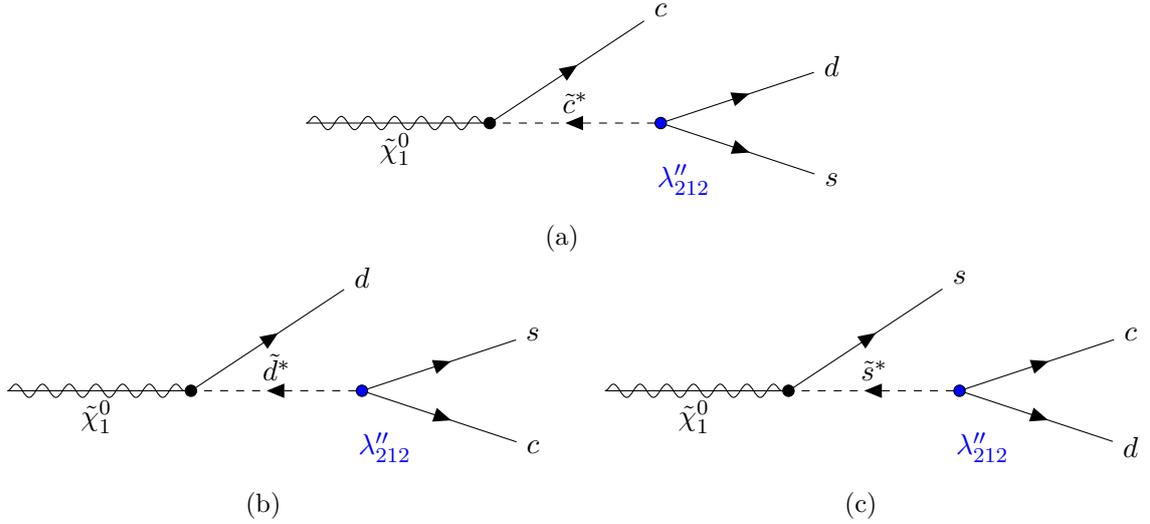
\begin{figure}[t]
    \centering

    \begin{minipage}{\textwidth}
    \centering
    \begin{subfigure}[b]{0.48\textwidth}
        \centering
        \begin{tikzpicture}[scale=1.5]
            \tikzfeynmanset{every vertex={dot, minimum size=2pt}}
            \begin{feynman}
                \vertex (v2) at (0, 0) {};
                \vertex [dot] (s) at (1.7, 0) {};
                \vertex (u) at (3.2, 1) {\(c\)};
                \vertex [dot, fill=blue] (v3) at (3.2, 0) {};
                \vertex (s2) at (4.7, 0.5) {\(d\)};
                \vertex (d) at (4.7, -0.5) {\(s\)};
    
                \diagram* {
                    (v2) -- [solid, edge label'=\(\tilde{\chi}^0_1\)] (s), 
                    (v2) -- [decorate, decoration={snake}, draw=black] (s), 
                    (s) -- [fermion] (u),
                    (v3) -- [scalar, dashed, with arrow=0.5, edge label'=\(\tilde c^*\)] (s),
                    (v3) -- [fermion] (s2),
                    (v3) -- [fermion] (d)
                };
    
                \node at (3.4, -0.5) {\(\textcolor{blue}{\lambda_{212}''}\)};
            \end{feynman}
        \end{tikzpicture}
        \caption{}
        \label{fig:diagram1}
    \end{subfigure}
    \end{minipage}

    \begin{minipage}{\textwidth}
    \centering
    \begin{subfigure}[b]{0.48\textwidth}
        \centering
        \begin{tikzpicture}[scale=1.5]
            \tikzfeynmanset{every vertex={dot, minimum size=2pt}}
            \begin{feynman}
                \vertex (v2) at (0, 0) {};
                \vertex [dot] (s) at (1.7, 0) {};
                \vertex (u) at (3.2, 1) {\(d\)};
                \vertex [dot, fill=blue] (v3) at (3.2, 0) {};
                \vertex (s2) at (4.7, 0.5) {\(s\)};
                \vertex (d) at (4.7, -0.5) {\(c\)};
    
                \diagram* {
                    (v2) -- [solid, edge label'=\(\tilde{\chi}^0_1\)] (s), 
                    (v2) -- [decorate, decoration={snake}, draw=black] (s), 
                    (s) -- [fermion] (u),
                    (v3) -- [scalar, dashed, with arrow=0.5, edge label'=\(\tilde d^*\)] (s),
                    (v3) -- [fermion] (s2),
                    (v3) -- [fermion] (d)
                };
    
                \node at (3.4, -0.5) {\(\textcolor{blue}{\lambda_{212}''}\)};
            \end{feynman}
        \end{tikzpicture}
        \caption{}
        \label{fig:diagram2}
    \end{subfigure}
    \hfill
    \begin{subfigure}[b]{0.48\textwidth}
        \centering
        \begin{tikzpicture}[scale=1.5]
            \tikzfeynmanset{every vertex={dot, minimum size=2pt}}
            \begin{feynman}
                \vertex (v2) at (0, 0) {};
                \vertex [dot] (s) at (1.7, 0) {};
                \vertex (u) at (3.2, 1) {\(s\)};
                \vertex [dot, fill=blue] (v3) at (3.2, 0) {};
                \vertex (s2) at (4.7, 0.5) {\(c\)};
                \vertex (d) at (4.7, -0.5) {\(d\)};
    
                \diagram* {
                    (v2) -- [solid, edge label'=\(\tilde{\chi}^0_1\)] (s), 
                    (v2) -- [decorate, decoration={snake}, draw=black] (s), 
                    (s) -- [fermion] (u),
                    (v3) -- [scalar, dashed, with arrow=0.5, edge label'=\(\tilde s^*\)] (s),
                    (v3) -- [fermion] (s2),
                    (v3) -- [fermion] (d)
                };
    
                \node at (3.4, -0.5) {\(\textcolor{blue}{\lambda_{212}''}\)};
            \end{feynman}
        \end{tikzpicture}
        \caption{}
        \label{fig:diagram3}
    \end{subfigure}
    \end{minipage}
    \caption{Parton-level diagrams for the decays of the neutralino to three quarks induced by the RPV SUSY coupling $\lambda_{212}''$.
    A set of diagrams for the charge-conjugated channels are implied.}
        \label{fig:decay-diagram}
\end{figure}

The Lagrangian can also lead to the neutralino decay $\neu \to {\cal B}M$ into a baryon and a meson, shown in Fig.~\ref{fig:decay-diagram}.
Given the mass hierarchy $m_B > m_\neu$ and the condition $\lambda''_{ijj}=0$, such decays can be mediated only by $\lambda''_{112}$ or $\lambda''_{212}$.
As discussed in Sec.~\ref{subsec:current-limits}, the current bounds on $\lambda''_{112}$ are much stronger.
Therefore, we consider only $\lambda''_{212}$.
The computation procedure for $\Gamma(\neu \to {\cal B}M)$  proceeds in a similar way to that of $\Gamma(B^+\to p \neu)$ and is elaborated on here.
For simplicity, we assume mass degeneracy of the relevant squark states, namely, $\tilde u_R$, $\tilde d_R$, $\tilde s_R$, $\tilde c_R$, and $\tilde b_R$.
We note that in this study the $\lambda''$ couplings are defined at the $B$-meson scale.

First, we define the invariant matrix element for the $\neu \to \mathcal{B} M$ decays as
\eq 
{\cal M} = \bar u_{\mathcal{B}}(p',s') \, 
\left[W_0^{LL}(p^2) - \frac{\not\! p}{m_{\neu}} W_1^{LL}(p^2)\right] 
\, P_L  \, u_{\neu}(p,s) \,, 
\en 
where $\bar u_{\mathcal{B}}(p',s')$ and $u_{\neu}(p,s)$ are the spinors of the final-state baryon and the initial-state neutralino, respectively; $p'$ and $p$ denote their momenta, with $m_{\mathcal{B}}$ and $m_{\neu}$ being their corresponding masses. We note that $q \equiv p-p'$ is thus the momentum of the meson $M$ with mass $m_M$. The functions  $W_0^{LL}(p^2)$ and $W_1^{LL}(p^2)$ are the transition form factors parameterizing the hadronic matrix elements
\eq 
\la \mathcal{B} M| {\mathcal O}^{LL} |0\ra = \bar u_{\mathcal{B}}(p',s') \, 
\left[ W_0^{LL}(p^2) - \frac{\not\! p}{m_{\neu}} W_1^{LL}(p^2)\right] 
\, P_L  \,,
\en
where the effective three-quark operator ${\mathcal O}^{LL}$ is obtained from the parton-level diagrams shown in Fig.~\ref{fig:decay-diagram} by neglecting the squark momentum in the propagator and amputating the neutralino field $\neu$. 
It is given by 
\eq 
{\mathcal O}^{LL} = 
  g^{\tilde cR} \, {\mathcal O}^{LL}_{cds}  
+ g^{\tilde dR} \, {\mathcal O}^{LL}_{dsc} 
+ g^{\tilde sR} \, {\mathcal O}^{LL}_{scd} 
\,, 
\en 
where $g^{\tilde qR} = g^{\tilde q}_{1R} \lambda''_{212}/m_{\tilde q}^2$ is the effective coupling of the neutralino to three quarks, 
${\mathcal O}^{LL}_{q_1q_2q_3} = 
\varepsilon_{a b c} \, 
\left(\bar q_{3}^c P_L C \bar q_{2}^b\right) 
\, \bar q_{1}^a P_L$, and $C=i \gamma^0 \gamma^2$ is the charge-conjugation matrix. 
The form factors $W_0^{LL}(p^2)$ and $W_1^{LL}(p^2)$ are calculated using the formalism discussed in detail in Ref.~\cite{Dib:2022ppx}, and their expressions are provided in Appendix~\ref{app:formfactors}.

The decay width of the neutralino into a baryon-meson pair, including the charge-conjugated mode, is given by 
\eq\label{decay_width}
& &\Gamma(\neu \to \mathcal{B}M + {\rm c.c.}) 
= \frac{\lambda^{1/2}(m_{\neu}^2,m_\mathcal{B}^2,m_M^2)}
{32 \pi m_{\neu}^3} 
\, \sum\limits_{\text{pol.}} \, |{\mathcal M}|^2 \,, \nonumber\\
& &\sum\limits_{\text{pol.}} \, |{\mathcal M}|^2 
= \left((m_{\neu}-m_\mathcal{B})^2 - m_M^2\right) \, (\mathcal{X}^2 + \mathcal{Y}^2)  
+ 2 m_\mathcal{B} m_{\neu}  \, (\mathcal{X} - \mathcal{Y})^2  
\,, 
\en 
where 
\eq 
           \mathcal{X} = W_0^{LL}(m_{\neu}^2)  
\,, \qquad \mathcal{Y} = W_1^{LL}(m_{\neu}^2)\,,
\en 
and 
\eq 
\lambda(x,y,z) = x^2 + y^2 + z^2 - 2 xy - 2 xz - 2yz 
\en 
is the kinematical K\"all\'en function.

\begin{table}[t]
\begin{center}
\vspace*{.25cm}
\def\arraystretch{1.1}
\begin{tabular}{|c|c|c|c|c|c|}
\hline 
\,\, Notation \,\,  & \,\, Content \,\, &  \,\, $J^P$ \,\, &
\,\, $S_d$ \,\, & \,\, Mass $m_{\cal B}$ (GeV)\,\, & \,\, 
Coupling  $\beta_{\cal B}$ (see Eq.~\eqref{eq:3quarkops})\,\, \\
 & & & & & in units of 10$^{-2} \times$ \!GeV$^3$
\\[2mm]
\hline
$\Lambda_{c}^+$  & $c[ud]$ & $1/2^+$ & $0$ & $2.28646$ & 0.835 \\
\hline
$\Sigma_{c}^0$  & $c\{dd\}$ & $1/2^+$ & $1$ & $2.45375$ & 1.125 \\
\hline
$\Xi_{c}^+$  & $c[us]$ & $1/2^+$ & $0$ & $2.46771$ & 1.021 \\
\hline
$\Xi_{c}^0$  & $c[ds]$ & $1/2^+$ & $0$ & $2.47044$ & 1.021 \\
\hline
$\Omega_{c}^{0}$ & $c\{ss\}$ & $1/2^+$ & $1$ & $2.6952$ & 2.325 \\
\hline
\end{tabular}
\end{center}
\caption{Classification of the relevant baryons, including their quantum numbers, masses, and coupling constants, extracted from Refs.~\cite{Faessler:2001mr,Gutsche:2019iac}.
}
\label{tab:baryons}
\end{table}

\begin{table}[t]
\begin{center}
\vspace*{.25cm}
\def\arraystretch{1.1}
\begin{tabular}{|c|c|c|c|c|}
\hline 
\,\, Notation \,\,  & \,\, Content \,\, &  \,\, $J^P$ \,\, &
\,\, Mass $m_M$ (GeV)\,\, & Decay constant  $f_M$ (GeV)\,\,
\\[2mm]
\hline
$\pi^0$  & $\frac{u \bar u - d \bar d}{\sqrt{2}}$ 
& $0^-$ & $0.1349768$ & $0.1302$ \\
\hline
$\pi^-$  & $d \bar u$ & $0^-$ & $0.13957$ & $0.1302$ \\
\hline
$K^-$  & $s \bar u$ & $0^-$ & $0.493677$ & $0.1557$ \\
\hline
$K^0$  & $d \bar s$ & $0^-$ & $0.497611$ & $0.1557$ \\
\hline
$\bar K^0$  & $s \bar d$ & $0^-$ & $0.497611$ & $0.1557$ \\
\hline
\end{tabular}
\end{center}
\caption{Classification of the relevant mesons, including their quantum numbers, masses, and leptonic decay constants, extracted from Ref.~\cite{Beringer:2024ady}.}
\label{tab:mesons}
\end{table}

Given the $\lambda''_{212}$ coupling, the neutralino decays into the quark-level final state $csd$. 
We consider five dominant hadronic final states and their charge-conjugated modes: 
\eq 
& &\neu \to \Sigma^0_c \bar K^0 + \mathrm{c.c.}\,, \quad 
   \neu \to \Omega^0_c K^0 + \mathrm{c.c.} \,, 
   \nonumber\\
& &\neu \to \Lambda_{c}^{+} K^- + \mathrm{c.c.} \,, \quad 
   \neu \to \Xi_{c}^{+} \pi^{-} + \mathrm{c.c.}\,, \quad
   \neu \to \Xi_{c}^{0} \pi^{0} + \mathrm{c.c.}
\label{eq:modes}
\en
In Table~\ref{tab:baryons} we present the properties of the final-state baryons, following Refs.~\cite{Faessler:2001mr,Gutsche:2019iac}.
These properties are the quark content, with $\{\cdots\}$ and $[\cdots]$ brackets indicating symmetric and antisymmetric spin configurations of the diquarks, respectively;
total angular momentum-parity $J^P$; total diquark spin $S_d$; mass $m_{\cal B}$; and coupling constant $\beta_{\cal B}$.
This coupling constant contains the matrix element of the three-quark operator ${\mathcal O}^{LL}$ between the corresponding baryon state and the vacuum, used in our calculation of the transition form factors and defined in Appendix~\ref{app:formfactors} (see also details in Ref.~\cite{Dib:2022ppx}).
The values of the coupling constants have been calculated using QCD sum-rule approaches~\cite{Azizi:2008ui,Azizi:2015bxa}.\footnote{See also Ref.~\cite{Domingo:2022emr} for a computation of a light neutralino decay widths in the low-mass regime.}

\begin{figure}[htp]
    \centering
    \includegraphics[width=.495\textwidth]{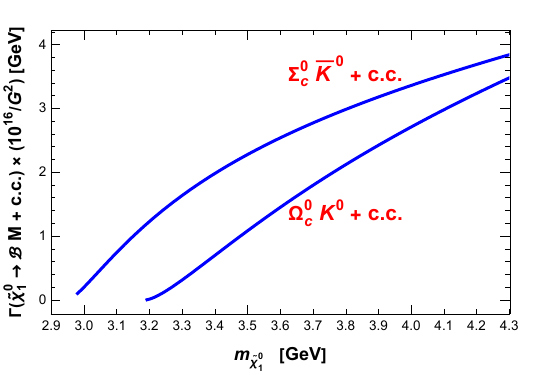}
    \includegraphics[width=.495\textwidth]{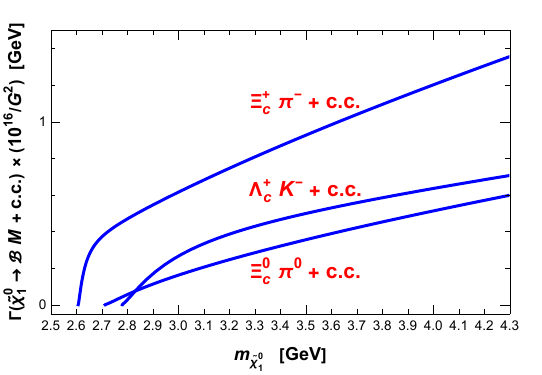}
\caption{Decay rates $\Gamma(\neu\to \mathcal{B}M)$, including the charge-conjugate mode, multiplied by $(10^{16}/G^2)$, where $G^2=({\lambda''_{212}})^2\times (1\text{ TeV}/m_{\tilde{q}})^4$.
}
\label{fig:neu_decay_width}
\end{figure}

\begin{figure}[htp]
\vspace*{-.2cm}
    \centering
    \includegraphics[width=.5\textwidth]{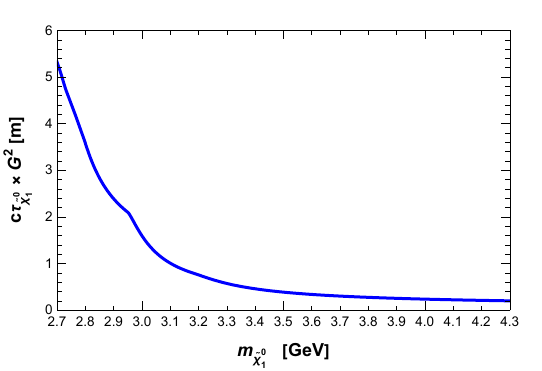}
\caption{The average proper decay length $c\tau_{\neu}$ of the neutralino, multiplied by $G^2=({\lambda''_{212}})^2\times (1\text{ TeV}/m_{\tilde{q}})^4$, as a function of $m_{\neu}$.
}
\label{fig:ctau}
\end{figure}

Table~\ref{tab:mesons} lists the properties of the final-state mesons, namely, the quark content, total angular momentum-parity $J^P$, mass $m_M$, and decay constants $f_M$, which are extracted from Ref.~\cite{Beringer:2024ady}.
The results of our calculation of the $\neu\to {\cal B}M +{\text c.c.}$ decay rates are shown in Fig.~\ref{fig:neu_decay_width} multiplied by the factor $10^{16}/G^2$, where $G^2=({\lambda''_{212}})^2\times (1\text{ TeV}/m_{\tilde{q}})^4$. 
Inverting the sum of these rates and multiplying by the speed of light $c$, we obtain the neutralino proper decay length $c\tau_{\neu}$, which is shown in Fig.~\ref{fig:ctau} multiplied by $G^2$, as a function of the neutralino mass.

We note that final states with a charmed meson and a strange baryon are suppressed relative to those in Eq.~(\ref{eq:modes}) as a result of the differences in the meson decay constants $f_M$ and baryon couplings $\beta_{\cal B}$.
Concretely, indicating these quantities by ($f_M^{\rm dom}$, $\beta^{\rm dom}$) and ($f_M^{\rm sup}$, $\beta^{\rm sup}$) for the dominant and suppressed modes, respectively, we find the ratio $\left(\dfrac{\beta^{\rm dom}}{\beta^{\rm sup}}\right)^2 \left(\dfrac{f_M^{\rm sup}}{f_M^{\rm dom}}\right)^2$ to be of the order of $\mathcal{O}(10)$. 
Therefore, we neglect these suppressed modes.
Final states involving a baryon and more than one meson are also possible.
However, na\"\i ve dimensional-analysis estimates of the two- and three-body decay rates of the neutralino into a baryon plus (one meson and two mesons), respectively, indicate suppression of the three-body decay mode by a factor of $1/(4 \pi)^2$, i.e., by about two orders of magnitude.
Thus, we simply estimate the branching fraction of the ignored modes with an inclusive calculation of the neutralino width at the parton level. 
For fully degenerate squarks, the inclusive width vanishes owing to a GIM-mechanism effect. 
However, removing the degeneracy by taking only one diagram in Fig.~\ref{fig:decay-diagram}, we obtain a total width similar to the sum of widths shown in Fig.~\ref{fig:neu_decay_width}.
Therefore, we conclude that the neglected modes do not significantly affect the total width and estimate our width calculation to be accurate up to uncertainties of order ~$\mathcal{O}(1)$.

\subsection{Current limits}
\label{subsec:current-limits}

Before proceeding to the experimental aspects of the proposed search, we discuss the current bounds on the RPV couplings of interest.

Bounds on BNV RPV couplings from searches at the Large Hadron Collider (LHC) are based on direct production of squarks.
Therefore, they are irrelevant for squark masses beyond the tightest bound, $m_{\tilde q}>1.85$~TeV~\cite{atlas:2020syg}.
This value was obtained in the channel $pp\to \tilde q\tilde q$ with $\tilde q\to q\neu$ under the assumption of 8 degenerate squark states and decoupled gluinos.

Limits are also derived from searches for dinucleon decays and baryon-antibaryon oscillations. 
Using the estimate of the baryon-antibaryon transition amplitude $\delta_{\mathcal B \overline{\mathcal{B}}}$ from Ref.~\cite{Aitken:2017wie},
\begin{equation}
    \delta_{\cal B \overline{\cal B}} \sim \frac{\kappa^2 m_{\chi_1^0}}{m_{\cal B}^2-m_{\chi_1^0}^2}\left(\frac{\lambda''_{ijk}\,g_{1R}^{\tilde q}}{m_{\tilde q}^2}\right)^2,\label{eq:delBB}
\end{equation}
where $\kappa\sim 0.01$ GeV$^{3}$~\cite{Buchoff:2015qwa}, we can translate the bounds on $\delta_{\cal B\overline{\cal B}}$ from Ref.~\cite{Aitken:2017wie} into bounds on the corresponding RPV couplings $\lambda''_{ijk}/m_{\tilde q}^2$.
In particular, the bounds $\delta_{(113)^2} \lesssim  10^{-13}\,\text{GeV}$, $\delta_{(112)^2} \lesssim 10^{-30}\,\text{GeV}$, and $\delta_{(212)^2} \lesssim 10^{-16}\,\text{GeV}$ derived from dinucleon decays (see Table I of Ref.~\cite{Aitken:2017wie}) translate into the limits shown in the third column of Table \ref{tab:existing_bounds}.

\begin{table}[h]
\centering
\begin{tabular}{|c|c|c|c|}
\hline
\multirow{2}{*}{$\lambda''_{ijk}$} & \multirow{2}{*}{$m_{\tilde{\chi}_1^0}\,\text{[GeV]}$} &  \multicolumn{2}{c|}{Bounds on $\lambda''_{ijk}/m_{\tilde q}^2\,\left[\text{TeV}^{-2}\right]$}
\\
\cline{3-4}
& & From Eq.~\eqref{eq:delBB} & From Eqs.~\eqref{eq:tauNNKK} and \eqref{eq:tauNNpipi} \\
\hline

\multirow{3}{*}{$\lambda''_{113}$}
& $2.7$ & $6\times 10^{2}$& $4\times 10^{3}$\\
\cline{2-4}
& $3.5$ & $5\times 10^{2}$ & $4\times 10^{3}$\\
\cline{2-4}
& $4.3$ & $3\times 10^{2}$ & $5\times 10^{3}$\\
\hline

\multirow{3}{*}{$\lambda''_{112}$}
& $2.7$ & $9\times 10^{-7}$& $5\times 10^{-7}$ \\
\cline{2-4}
& $3.5$ & $1\times 10^{-6}$ & $5\times 10^{-7}$\\
\cline{2-4}
& $4.3$ & $1\times 10^{-6}$ & $6\times 10^{-7}$\\
\hline

\multirow{3}{*}{$\lambda''_{212}$}
& $2.7$ & $4\times 10^{0}$& $1\times 10^{2}$ \\
\cline{2-4}
& $3.5$ & $8\times 10^{0}$ & $1\times 10^{2}$\\
\cline{2-4}
& $4.3$ & $1\times 10^{1}$ & $1\times 10^{2}$\\
\hline

\end{tabular}
\caption{Estimates of existing bounds on the RPV couplings for selected neutralino masses.}\label{tab:existing_bounds}
\end{table}

Another way to estimate the existing bounds from dinucleon decay is to use the expression for the nuclear-matter lifetime given in Eq.~(4.3) of Ref.~\cite{Calibbi:2016ukt}, after applying the replacements $\alpha_S C_F \to \alpha_2 \tan^2 \theta_W e_q^2$ and $m_{\tilde g}\to m_{\tilde \chi_1^0}$:\footnote{Eq.~(4.3) of Ref.~\cite{Calibbi:2016ukt} assumes the process is mediated by gluino exchange between nucleons. In our case, the gluino is decoupled from the spectrum, and the process instead proceeds via the exchange of a virtual neutralino.}
\begin{align}
    \tau_{NN\to KK} = \frac{2 m_N^2 m_{\tilde \chi_1^0}^2 m_{\tilde s_R}^8}{9\pi \alpha_2^2 (\lambda''_{112})^4 e_s^4 \tan^4\theta_W \rho_N\left<KK|(u_R d_R s_R)^2|NN\right>^2},\label{eq:tauNNKK}
\end{align}
where $\rho_N = 0.25\,\text{fm}^{-3}$ is the nuclear matter density, $\alpha_2=g_2^2/4\pi$, $e_s=-1/3$, and $m_N$ is the nucleon mass. The hadronic matrix element is assumed to be $\left<KK|(u_R d_R s_R)^2|NN\right>\approx (150\,\text{MeV})^5$.
Using the experimental bound from Super-Kamiokande~\cite{Super-Kamiokande:2014hie}, $\tau_{pp\to K^+K^+}>1.7\times 10^{32}$ years, we derive the corresponding bounds on $\lambda''_{112}$, shown in Table~\ref{tab:existing_bounds}.

In principle, a similar expression could be used to constrain $\lambda''_{111}/m^2_{\tilde{q}}$, if this coupling were non-zero, through the process $NN\to \pi\pi$, with
\begin{align}
    \tau_{NN\to \pi\pi} = \frac{2 m_N^2 m_{\tilde \chi_1^0}^2 m_{\tilde d_R}^8}{9\pi \alpha_2^2 (\lambda''_{111})^4 e_d^4 \tan^4\theta_W \rho_N\left<\pi\pi|(u_R d_R d_R)^2|NN\right>^2}\, ,\label{eq:tauNNpipi}
\end{align}
by considering, e.g., the Super-Kamiokande limit $\tau_{nn\to\pi^0\pi^0} > 4.04 \times  10^{32}\,\text{years}$~\cite{Super-Kamiokande:2015jbb}.
Although $\lambda''_{111}$ vanishes identically, Eq.~\eqref{eq:tauNNpipi} can be used to derive bounds on $\lambda''_{212}/m^2_{\tilde{q}}$ and $\lambda''_{113}/m^2_{\tilde{q}}$ by adding two weak insertions to the diagram, which reduce the amplitude by a factor of approximately 
$\left[\pi V_{us} V_{cd}^* \, G_F f_\pi^3 /m_c\right]^2
\sim 10^{-17}$ and $\left[\pi V_{ub} V_{ud}^* \, G_F  f_\pi^3 /m_b\right]^2 \sim 10^{-20}$, respectively, 
where $V_{q_1q_2}$ are the CKM matrix elements, $G_F$ is the Fermi constant, and
$m_{c/b}$ is the mass of the charm/bottom quark. 
The resulting bounds are presented in the rightmost column of Table~\ref{tab:existing_bounds} for benchmark values of the neutralino mass.

In addition, the BABAR collaboration~\cite{BaBar:2023dtq} has searched for the decay $B^+\to p\neu$ where the neutralino escapes the detector, and interpreted the results in the scenario, proposed in Ref.~\cite{Dib:2022ppx}, of a light neutralino and $\lambda''_{113}$ being the only nonzero RPV coupling.
Using a data sample with an integrated luminosity of 398~fb$^{-1}$, BABAR obtained $m_{\neu}$-dependent upper limits on $\lambda''_{113}/m^2_{\tilde{q}}$ between about $0.1$ and $1$~TeV$^{-2}$ for $m_{\neu}=2.7$ and $4.3$~GeV, respectively. 
It should be noted that these bounds weaken with increasing values of $\lambda''_{212}/m^2_{\tilde{q}}$, as more neutralino decays take place inside the detector and are rejected by the missing-energy requirement applied in the BABAR analysis.

Concerning the bounds on the couplings $\lambda''$, one must consider that their derivation should obey perturbative unitarity, which is valid for small couplings.
Thus, for large values of $\lambda''$, there may still be experimental bounds, but perturbative calculations may not be reliable for extracting those bounds.

\section{Experimental techniques}
\label{sec:techniqueANDbackground}

We propose to conduct the search at the Belle~II experiment, which is described briefly in Sec.~\ref{subsec:belleii}.
We consider two approaches, inclusive and exclusive, explained in detail below in Secs.~\ref{subsec:inclusive} and~\ref{subsec:exclusive}, respectively. 
The inclusive approach is novel and is proposed here for the first time. 
Its main advantage is utilization of essentially all the decay modes of the neutralino.
While in principle it suffers from higher background, we discuss in Sec.~\ref{subsubsec:bgd-supp} methods for background suppression, arguing that it can be kept at the 1-event level for a given value of neutralino mass.
In Sec.~\ref{subsubsec:efficiency} we estimate the signal-reconstruction efficiencies for both approaches.

\subsection{The Belle~II Experiment}
\label{subsec:belleii}

The Belle~II detector~\cite{Belle-II:2010dht,Belle-II:2018jsg} collects data at the SuperKEKB~\cite{Ohnishi:2013fma,Akai:2018mbz} collider at the KEK laboratory in Tsukuba, Japan.
SuperKEKB collides electron and positron beams with energies of 7 GeV and 4 GeV, respectively.
The resulting center-of-mass (CM) energy $\sqrt{s}=10.58$~Gev corresponds to the mass of the $\Upsilon(4S)$ resonance, which decays promptly to two $B$-mesons.
Belle~II is a magnetic spectrometer of cylindrical geometry placed around the SuperKEKB interaction point (IP). 
The detector consists of several subsystems.
The subsystem closest to the IP is a vertex detector (VXD) composed of two layers of silicon pixel detectors and four layers of silicon strip detectors.
Outside the vertex detectors is a helium-based central drift chamber (CDC). 
The combined VXD-CDC system measures  charged-particle momenta in a $1.5$~T magnetic field provided by a superconducting solenoid.
Charged hadrons are identified mainly by end-cap and barrel Cherenkov devices located outside the CDC, with additional information from specific ionization in the CDC.
Belle~II has not yet published the proton identification performance of the detector.
However, it is expected to be at least as good as that of the BABAR detector's Cherenkov subsystem. 
For a proton identification efficiency of 90\% or more, BABAR reports less than 2\% pion efficiency across the relevant momentum range~\cite{BaBarDIRC:2004tub}.
Outside of the Cherenkov subsystems, Belle~II has an electromagnetic calorimeter and a muon- and~$K_L^0$-identification system.

\subsection{Inclusive search}
\label{subsec:inclusive}

\subsubsection{Partial reconstruction}
\label{subsubsec:partial}

The search channel involves the decay $B^+\to p \neu$. 
We study the case in which the neutralino is long-lived but decays within the detector tracking volume. 
Since in this study the neutralino is relatively heavy, between $2.7$ and $4.3$~GeV, it can decay via a large number of decay chains, each having a small branching fraction.
In addition, many decay chains lead to final states that contain $\pi^0$ mesons, for which the reconstruction efficiency is lower and the background is higher than for charged-particle tracks.

To overcome this limitation, we suggest to only partially reconstruct the neutralino.
In this new approach, one needs only to reconstruct the DV produced in the neutralino decay. 
Thus, the minimal number of observed displaced tracks is 2, although we require 3 tracks to ensure sufficient background suppression, as discussed in Section~\ref{subsubsec:bgd-supp}. 
We note that the neutralino final state may include additional particles. 
These may be charged particles outside the angular coverage of the tracker or that the reconstruction algorithm failed to detect, photons produced mainly in $\pi^0$ decays, and the detector-stable neutron and $K_L^0$ meson.
We note that, owing to the baryonic final state, a neutron is produced in about half of the neutralino decays.
Since partial reconstruction requires only 3 tracks, it maintains high efficiency even in the presence of such unreconstructed particles, constituting the main advantage of this approach. 
Its disadvantage is a higher background than in the case of full reconstruction. 
However, in Sec.~\ref{subsubsec:bgd-supp} we argue that, with proper event selection, the background level is expected to be very low.

Despite the unreconstructed particles, we show that the invariant mass of each neutralino candidate can be measured by applying the constraints of the decay. 
This enables background suppression and provides the neutralino mass if a signal is observed.
To see this, we note that, since the neutralino is not fully reconstructed, there are 8 unknown kinematic quantities: the four-momenta of the neutralino and the $B^+$.
There are also 8 constraints. 
Four constraints arise from four-momentum conservation in the $B^+$ decay. 
Two constraints arise from equality between the direction $\hat p_\neu$ of the neutralino momentum and the vector connecting the $e^+e^-$ interaction point to the measured position of the DV, neglecting the small $B^+$ lifetime.\footnote{In the experimental analysis, the production point of the neutralino can be determined by a vertex fit that constrains the proton to originate from the beamspot -- the spatial region that contains 68.3\% of the collision points, expanded to account for the $B$-meson lifetime. 
We do not carry out this procedure, leading to somewhat increased smearing in the calculation of the kinematic quantities.} 
One constraint corresponds to the known $B^+$ mass. 
Finally, the energy $E_B$ in the CM frame of the $e^+e^-$ collision must equal half the CM energy, $\sqrt{s}/2$. 
The value of $\sqrt{s}/2$ and the boost vector $\vec\beta$ between the laboratory frames are known with small uncertainties.

The application of the constraints and calculation of all the kinematic quantities of the decay are detailed in Appendix~\ref{app:mass}. 
This calculation involves a quadratic equation, giving rise to two solutions.
In order for the calculation to be valid, we require the equation's discriminant $D$ to be non-negative.
For these cases, one obtains two solutions for the neutralino mass.
We label these $m_+$, $m_-$, with the superscript referring to the sign chosen for $\pm \sqrt{D}$. 
In a given signal event, one of these sign choices is correct, resulting in either $m_+$ or $m_-$ being close to the true neutralino mass $m_\neu$ to within the resolution. 
By contrast, this is not the case for background events. 
Therefore, $m_+$ and $m_-$ provide a way to suppress the background,  as discussed in Sec.~\ref{subsubsec:bgd-supp}.
Furthermore, in the case of an observed signal, these variables provide a measurement of the neutralino mass.

\subsubsection{Background suppression}
\label{subsubsec:bgd-supp}

In this section we consider background processes that can fake the experimental signature, which is a DV and a proton that is promptly produced (neglecting the small $B^+$ lifetime).
To ensure that the DV is significantly displaced from the IP, we require its radial position to satisfy $r_\DV>1$~cm, which is many times the detector resolution~\cite{TrackRes}.

Background arises from both $e^+e^-\to B\bar B$ or $e^+e^-\to q\bar q$  processes, where $q$ indicates an up, down, strange, or charm quark.
These processes give rise to many promptly produced tracks, as well as displaced tracks produced in decays of long-lived strange particles or in material interactions.  
In what follows we first consider sources of 2-track DVs. 
Although most of them can be efficiently removed with kinematic cuts, we argue that enough remain to constitute non-negligible background.
In an actual data analysis, the level of background can be more accurately studied and potentially mitigated.
Within the limited scope of the current study, we opt for strongly suppressing all such background by requiring DVs to have at least 3 tracks with invariant mass $m_\DV>1.5$~GeV.

The first source of 2-track background DVs is the decays of the long-lived hadrons $K^0_S\to \pi^+\pi^-$ and $\Lambda\to p\pi^-$. 
These decays can be easily suppressed by vetoing 2-track DVs with invariant mass $m_{\DV}$ close to the known masses of the $K^0_S$ and $\Lambda$.
For example, in Ref.~\cite{Belle:2024wyk}, Belle suppressed $K^0_S$ background with the broad cut $m_{\DV} \notin [420, 520]$~MeV.
Further suppression can be obtained by requiring a maximal value for $\hat p_\DV \cdot \hat r_\DV$, the cosine of the angle between the momentum vector of the tracks emanating from the DV and the vector connecting the IP to the DV.
This utilizes the fact that neutralino decays give rise to additional particles beyond the two constituting the DV. 
As a counterexample, BABAR has used $\hat p_\DV \cdot \hat r_\DV > 0.999$ for the opposite purpose of selecting $K_S^0$ decays~\cite{BaBar:2012iuj}.

The second source of 2-track background DVs is decays of the $K^0_L$ meson.
This background is suppressed by the long average proper decay length of the $K^0_L$, $c\tau_{K^0_L}\approx 15$~m.
However, since the $K_L^0$ undergoes a 3-body decay to $\pi^+\pi^-\pi^0$ and $\ell^\pm\pi^\mp\nu$, $m_{\DV}$ can be as small as the kinematic threshold, necessitating rejection of all DVs  with $m_{\DV} < 520$~MeV.

The third source is interaction of particles with detector material. 
This includes photon conversions to $e^+e^-$ and hadronic interactions that produce mesons and potentially protons or nuclear fragments ejected from the material.
Such DVs are suppressed by the $m_{\DV}$ cut and can be further suppressed by requiring that the DV is not within or near dense detector material.
For example, in Ref.~\cite{Belle:2024wyk}, Belle applied the aggressive requirement $r_{\DV} > 15$~cm on the radial position of the DV, placing the DV in the gaseous volume of the CDC and beyond the solid material of the VXD.
This also helps suppress background from $K^0_S$ and $\Lambda$ decays. 
This cut reduces the signal efficiency, but mostly at lower lifetimes that are not at the edge of the experimental sensitivity.
Photon conversions are further suppressed by rejecting low-$m_\DV$ DVs.

The fourth source of 2-track background DVs is coincidental crossings of tracks.
Since most tracks are prompt, i.e., produced close to the IP, this background is greatly suppressed by requiring that DV tracks have a minimal impact parameter with respect to the IP.
This requirement benefits from the small SuperKEKB beamspot size, which is a few microns in the horizontal dimension and less than $100$~nm in the vertical dimension~\cite{Belle-II:2010dht}.
Furthermore, the resolution of a track's transverse impact parameter is less than $50~\mu$m for transverse momentum $p_{\text{T}}>500$~MeV, and remains sub-millimeter even for very soft tracks with $p_{\text{T}}=100$~MeV~\cite{BelleIITrackingGroup:2020hpx}.
Additional suppression of coincidental-crossing  background is achieved by requiring that each DV track should not have detector hits on both sides of the DV.
This also suppresses background from tracks that originate from interactions of off-orbit beam particles with material.
The impact-parameter requirement does not suppress background involving displaced tracks. 
These can be produced in decays of LLPs such as $K_S^0$, $\Lambda$, as well as of charged pions and kaons.
They can also be prompt tracks that underwent hard scattering in detector material or that were misreconstructed owing to close proximity to other tracks, losing their hit association as a result.

Despite the application of such background-suppression measures, in Ref.~\cite{Belle:2024wyk} Belle found 273 di-hadron DVs with $0.520 < m_{\DV} < 1.638$~GeV (calculated with the pion mass hypothesis for both tracks) in a data sample with an integrated luminosity of $915~{\rm fb}^{-1}$.
Simulation studies showed that most of those DVs were formed by tracks from $K_S^0$ or $\Lambda$ decays.
Since the di-hadron DVs constituted only a validation sample in Ref.~\cite{Belle:2024wyk}, no further attempt was made to reduce the background, and it is not clear what fraction would survive the mass requirement $m_\DV>1.5$~GeV.

We conclude that within the actual data analysis, 2-track DVs may be usable as a valid signal channel following more detailed studies that demonstrate sufficient background suppression.
However, here we restrict ourselves to 3-track DVs, guaranteeing strong background suppression and simplifying our study.

A potential background source for 3-track DVs is hadronic material interactions.
Since these are much rarer than 2-track DVs, for which we have a rough estimate from Ref.~\cite{Belle:2024wyk}, we conclude that they can be pushed to negligible levels, particularly with the requirement $m_\DV>1.5$~GeV.
Another potential source is coincidental crossing of a third track with a 2-track DV, itself arising from any of the sources discussed above.  
Since the 2-track DV is a pointlike object, the probability of crossing it is much smaller than that for coincidental crossing with another track, which is a 1-dimensional object.
As a result, the number of 3-track coincidental-crossing DVs is much smaller than that of 2-track DVs.
This background is further strongly suppressed by vetoing DVs for which two tracks are a $\pi^+\pi^-$ or $p\pi^-$ pair with mass close to that of the $K_S^0$ or $\Lambda$, respectively.
Thus the expected 3-track-DV background yield is much smaller than the $273$ events per $915~{\rm fb}^{-1}$ found in Ref.~\cite{Belle:2024wyk}.
We refer to this yield as the baseline background.

Next, we estimate the additional background suppression provided by the partial-reconstruction procedure described in Sec.~\ref{subsubsec:partial}. 
We use the EVTGEN MC generator~\cite{evtgen} to generate samples of signal events with the neutralino decaying to the five decay modes of Eq.~\eqref{eq:modes} with the branching fractions corresponding to the decay widths shown in Fig.~\ref{fig:neu_decay_width}.
Signal samples are generated for $m_\neu=2.7$~GeV to $m_\neu=4.3$~GeV in $100$-MeV-wide steps.
For background simulation, we use a simplified model to mimic the kinematic behavior: we produce $e^+e^-\to B\bar B$ events with EVTGEN and $e^+e^-\to s\bar s$ events with Pythia~8~\cite{Sjostrand:2014zea}, with all particles decaying generically according to the default models in the generators.
The background is partially reconstructed by selecting a proton and a $K_S^0$, taking the DV position to be the decay position of the $K_S^0$.
Requiring $D\ge 0$ retains between 80\% and 73\% of the signal events, depending on $m_\neu$, and removes 94\% of the $B\bar B$ background and 91\% of the $s\bar s$ background.

\begin{figure}
    \centering
        \begin{subfigure}{0.45\textwidth}
            \centering
            \includegraphics[width=\textwidth]{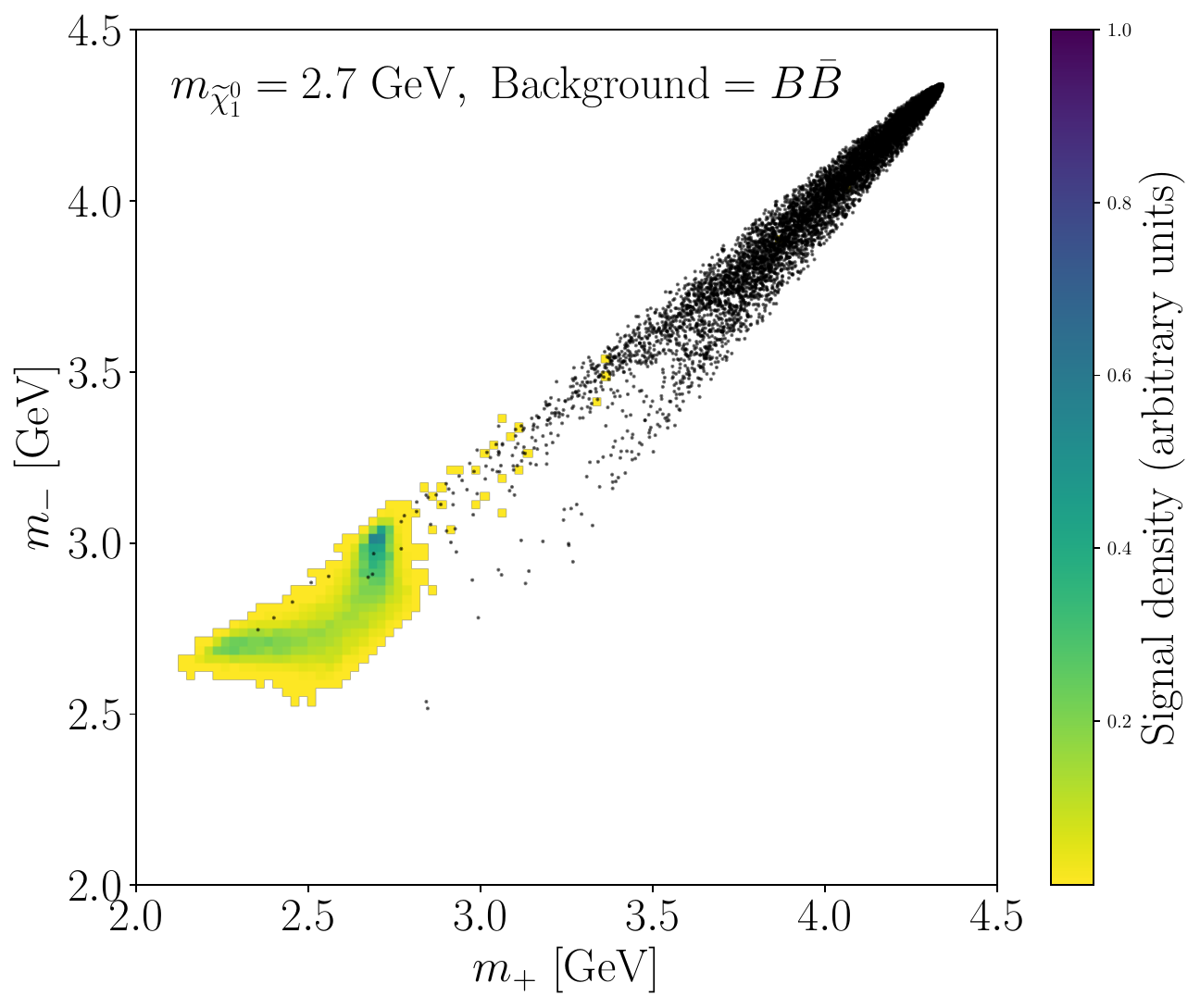}
            \caption{}
            \label{subfig:2.7-BB}
        \end{subfigure}
        \begin{subfigure}{0.45\textwidth}
            \centering            
            \includegraphics[width=\textwidth]{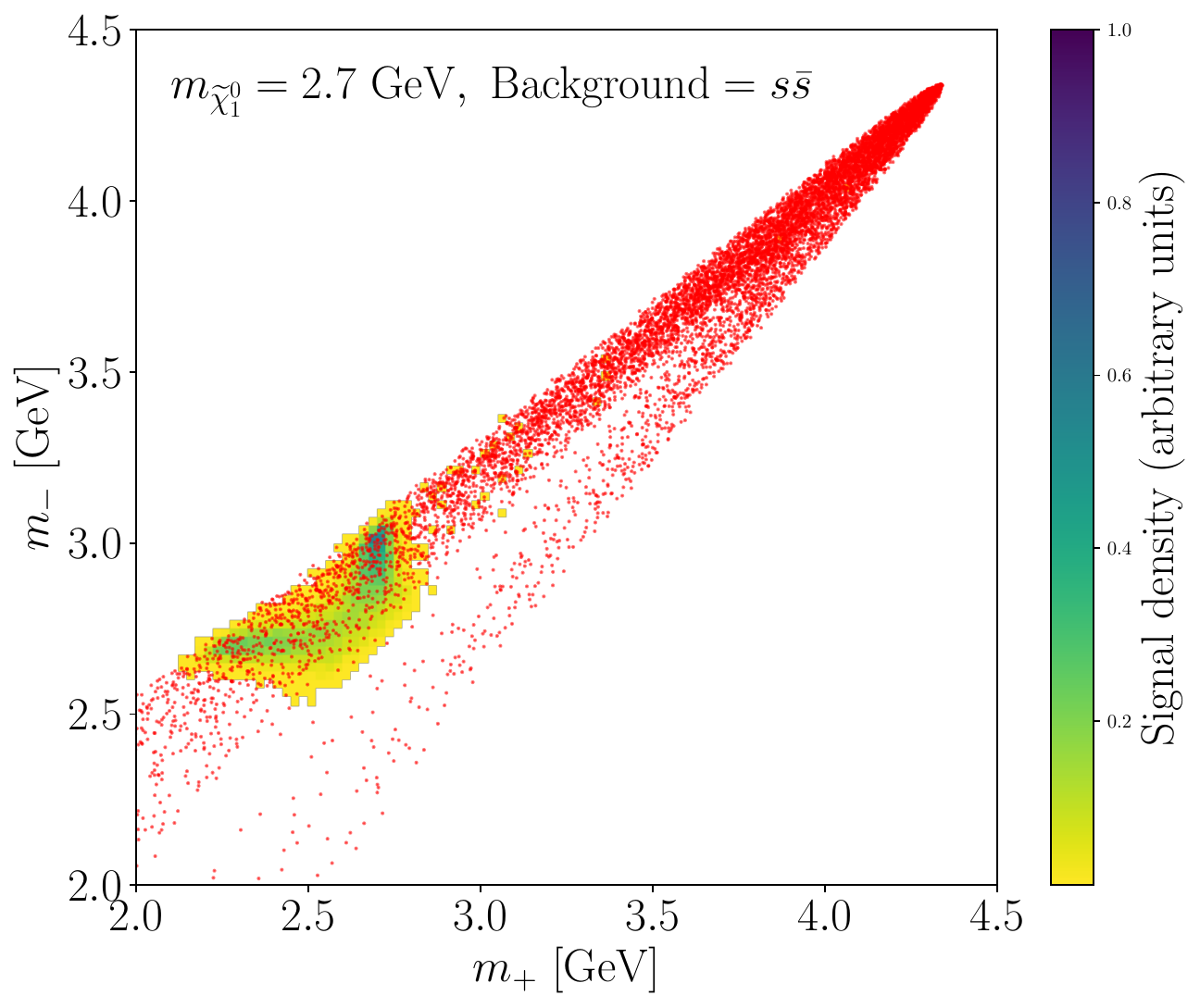} 
            \caption{}
            \label{subfig:2.7-ss}
        \end{subfigure}
    \\
        \begin{subfigure}{0.45\textwidth}
            \centering
            \includegraphics[width=\textwidth]{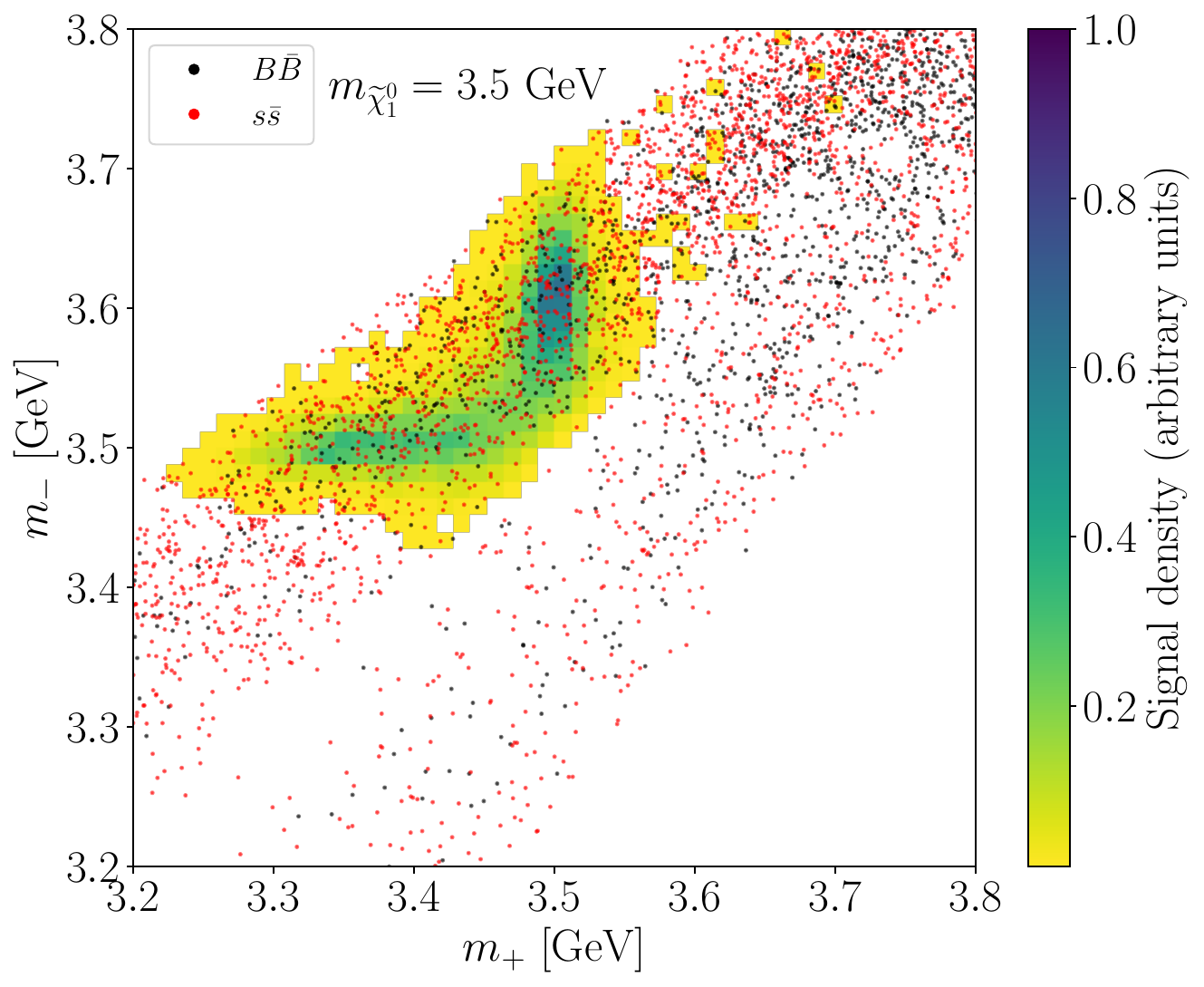}
            \caption{}
            \label{subfig:3.5}
        \end{subfigure}
        \begin{subfigure}{0.45\textwidth}
            \centering            
            \includegraphics[width=\textwidth]{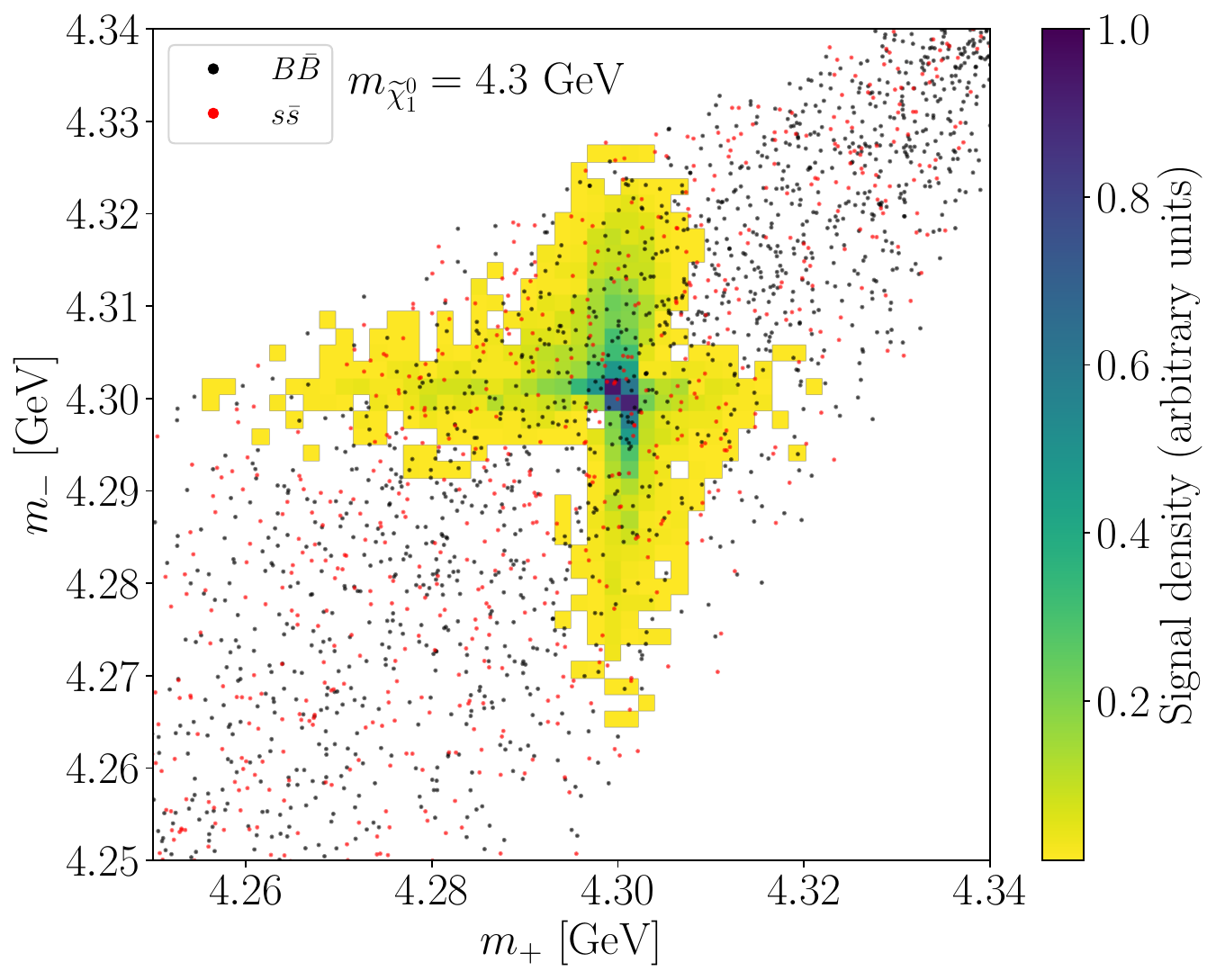} 
            \caption{}
            \label{subfig:4.3}
        \end{subfigure}
   \caption{The distributions of $m_-$ vs. $m_+$ for signal events (colored bins) and for $e^+e^-\to B\bar B$ (black dots) and $e^+e^-\to s\bar s$ (red dots) background-model events.
   The two upper panels show the $m_\neu=2.7$~GeV signal with the $B\bar B$ and $s\bar s$ background events, respectively, in the full $m_+$ and $m_-$ range.
   The lower panels zoom in on the $m_\neu=3.5$~GeV and $m_\neu=4.3$~GeV signals, respectively, overlaid with both $B\bar B$ and $s\bar s$ background events.
   \label{fig:m+m-}}
\end{figure}

The resulting $m_\pm$ distributions for signal and background are shown in Fig.~\ref{fig:m+m-}.
The spread in the signal distributions reflects the spread in the collider center-of-mass energy $\sqrt{s}$.
For each value of $m_\neu$, we select the $m_+$-vs.-$m_-$ region that contains 90\% of the signal events. 
We observe that this retains between 0.5\% and 7\% of the background, depending on $m_\neu$ and the type of background.
Thus, we estimate that applying these selections retains at most 9\% of the baseline background for all values of $m_\pm$, and at most 0.6\% for any given hypothesis on $m_\neu$.

We note that, if needed, additional cuts can be used to further suppress background. 
First, requiring that the DV have 4 tracks instead of 3 should strongly suppress background for the reasons outlined above for the 3-track requirement. 
This will reduce the efficiency by a few ten percent, as discussed in Sec.~\ref{subsubsec:efficiency}.
Additionally, background suppression to the level of a few percent can be obtained by exploiting the Majorana nature of the neutralino and the proton-identification capabilities of the detector.
Specifically, one requires the presence of a proton emanating from the DV, with the same charge as the prompt proton emitted in the $B^+\to \neu p$ decay.
This discards about $75\%$ of the signal, since protons are produced in half the decays, and half of those have the opposite charge as the prompt proton.
Additional loss of efficiency arises from the requirement that the proton should be detected and identified.
Further limiting the sample to only $B^-\to \neu \bar{p}$ decays can be used to suppress protons emitted from material interactions, at the cost of another 50\% of the signal.
However, we estimate that these measures will not be needed and do not account for them in our estimates.

\subsubsection{Methods for background estimation}
\label{subsubsec:bgd-est}

Even in cases in which the expected background level is less than one event, one still needs a way to estimate it and to assign uncertainties to that estimate. 
Although this is part of the eventual data analysis, we briefly note on some methods with which it can be done.

Background from material interactions can be studied from DVs situated inside dense material regions, such as the beampipe wall.

Background from coincidental crossing can be estimated by allowing one or more of the DV tracks to have detector hits on both sides of the DV. 
It can be further studied by forming a control sample of 3-track DVs, each composed of a 2-track DV from one event and third track from a different event. 
This event-mixing method has been used, e.g., in Ref.~\cite{ATLAS:2015oan} by the ATLAS Collaboration.
The procedure results in a very large sample of coincidental-crossing different-event DVs, from which the crossing probability can be extracted with high statistical precision. 
In general, the crossing probability depends on potential correlations between the physical origin of the 2-track DV and that of the third track, which do not exist for different-event DVs.
This can be accounted for by computing the crossing probability as a function of variables such as the azimuthal and polar angle between the track momentum and $\hat r_\DV$.

Finally, this event-mixing method can also be used to combine a DV from one event with a prompt proton from another event, similar to an approach taken by ATLAS in Ref.~\cite{ATLAS:2022atq}.
This yields a high-statistics sample from which to estimate the probability for observing both the prompt proton and the DV in the same event.
In fact, depending on the nature of the background observed, this simple approach might be sufficient for estimating all sources of background.

\subsection{Exclusive search}
\label{subsec:exclusive}

The inclusive, displaced search described in Sec.~\ref{subsec:inclusive} uses cuts on $r_\DV$, $D$, and $m_\pm$ to suppress background and maintain high efficiency for almost all neutralino decay chains.
An alternative search strategy is an exclusive search, in which one fully reconstructs only a few  final states that have an optimal combination of branching fraction and efficiency. 
This approach removes the two-fold ambiguity in the calculation of $m_\neu$ and achieves additional background reduction through the use of the well-known kinematic quantities $\Delta E = E_B - \sqrt{s}/2$ and 
$M_{bc}=\sqrt{s/4 - (p_B)^2}$, where $p_B$ and $E_B$ are the CM-frame momentum and energy of the fully reconstructed $B^+$.
Exploiting the neutralino lifetime, the requirement $r_{\DV}>1$~cm must also be used to suppress the background to negligible levels.

The main disadvantage of this method is the small branching fraction for any specific final state.
Experimentally, the best decay chains are those in which the neutralino decays to 4 or 6 tracks and no neutral particles.
We consider the dominant decay chains into these final states, taking their branching fractions from Ref.~\cite{Beringer:2024ady}.
\begin{itemize}
\item For $m_\neu < m_{\Lambda_c^+} + m_{K^-} = 2.784$~GeV, only the decay $\neu\to \Xi_c^+\pi^-$ is kinematically allowed. 
The dominant $\Xi_c^+$ decay chain into an easily reconstructible final state is $\Xi_c^+\to \Xi^-\pi^+\pi^+$ with $\Xi^-\to \Lambda \pi^-$ and $\Lambda\to p\pi^-$, resulting in 6 tracks. 
The total branching fraction for this decay chain is $1.9\%$, excluding the neutralino decay.
We take this at face value, ignoring the large relative uncertainty of $45\%$ on the branching fraction $\text{B}(\Xi_c^+\to \Xi^-\pi^+\pi^+)$.

\item For $m_\neu > m_{\Sigma_c^0} +m_{\bar K^0} = 2.951$~GeV, the dominant decay chain that culminates in a 6-track final is $\neu\to \Sigma_c^0 \bar K^0$, with $\Sigma_c^0\to \Lambda_c^+ \pi^-$ and $\Lambda_c^+\to p K^- \pi^+$.
The $\bar K^0$ must decay as $K^0_S \to \pi^+\pi^-$ to be observable, so that the final state has 6 tracks. 
The total branching fraction, excluding that of the $\neu$ decay, is 2.2\%.

\item The 4-track final state is dominated by the decay $\neu\to \Lambda_c^+ K^-$ followed by $\Lambda_c^+ \to pK^- \pi^+$, which has a branching fraction of 6.4\% and is kinematically allowed for $m_\neu > m_{\Lambda_c^+} + m_{K^-} = 2.782$~GeV.
\end{itemize}

We note that for $m_\neu > m_{\Omega_c^0} + m_{K^0} = 3.192$~GeV, the channel $\neu\to \Omega_c^0 K^0$ opens as well, followed by $\Omega_c^0 \to \Omega^- \pi^+$, $\Omega^-\to \Lambda K^-$, and $\Lambda\to p\pi^-$, with $K^0\to \pi^+\pi^-$.
The branching fraction for $\Omega_c^0 \to \Omega^- \pi^+$ has not been measured, but is expected to be 3.4\%~\cite{Zeng:2024yiv}. 
Thus, the total branching fraction for this decay chain, excluding the neutralino decay, is 0.5\%.
Accounting also for the fact that $\text{B}(\neu\to \Omega_c^0 K^0) < \text{B}(\neu\to \Sigma_c^0 \bar K^0)$ (see Fig.~\ref{fig:neu_decay_width}), we conclude that $\neu\to \Omega_c^0 K^0$ offers no significant advantage, and we do not take it into account in our estimates.

It is worth mentioning that the exclusive technique can also be used without requiring a DV, focusing on short neutralino lifetimes. 
In this case, there is background from the SM decays $B^+\to p \bar\Xi_c^+ \pi^-$, $B^+\to p \bar\Omega_c^0 \bar{K}^0$, and $B^+\to p \bar\Lambda_c^- K^+$.
These are Cabibbo-suppressed $\bar b\to \bar c u \bar s$ transitions that have not yet been studied~\cite{Beringer:2024ady}, and hence are also of interest beyond the topic of this paper.
Such background can be suppressed by requiring the final-state baryon number to be $\pm 2$, at a cost of losing half the signal.
Since short neutralino lifetimes correspond to relatively large values of $\lambda''_{212}$, this approach is less favorable.
Nonetheless, it may be suitable as an early search that also explores for the first time these Cabibbo-suppressed SM decay modes.

\subsection{Tracking-efficiency estimation}
\label{subsubsec:efficiency}

Reconstruction of a signal event requires reconstructing the proton and the tracks that make up the DV. 
These are at least 3 tracks for the inclusive search, and exactly 4 or 6 tracks for the exclusive search.
The displaced-track reconstruction efficiency depends on the position of the neutralino decay, the number of charged particles produced in the decay, and their momentum vectors. 
Accurate determination of the reconstruction efficiency requires full detector simulation of a large event sample and validation samples from detector data. 
Such samples are available only for experimental studies performed and published by the experimental collaboration.
Therefore, we use a geometric estimation of the track reconstruction efficiency using the TrackEff package~\cite{Bertholet:2025lcr,TrackEff}.
Each track in our simulated signal events is passed through TrackEff to estimate the number of CDC hits on the track.
Following the recommendation in Ref.~\cite{Bertholet:2025lcr}, we consider a track as reconstructed if it has at least 20 hits.

\begin{figure}
\centering
\includegraphics[width=1.0\textwidth]{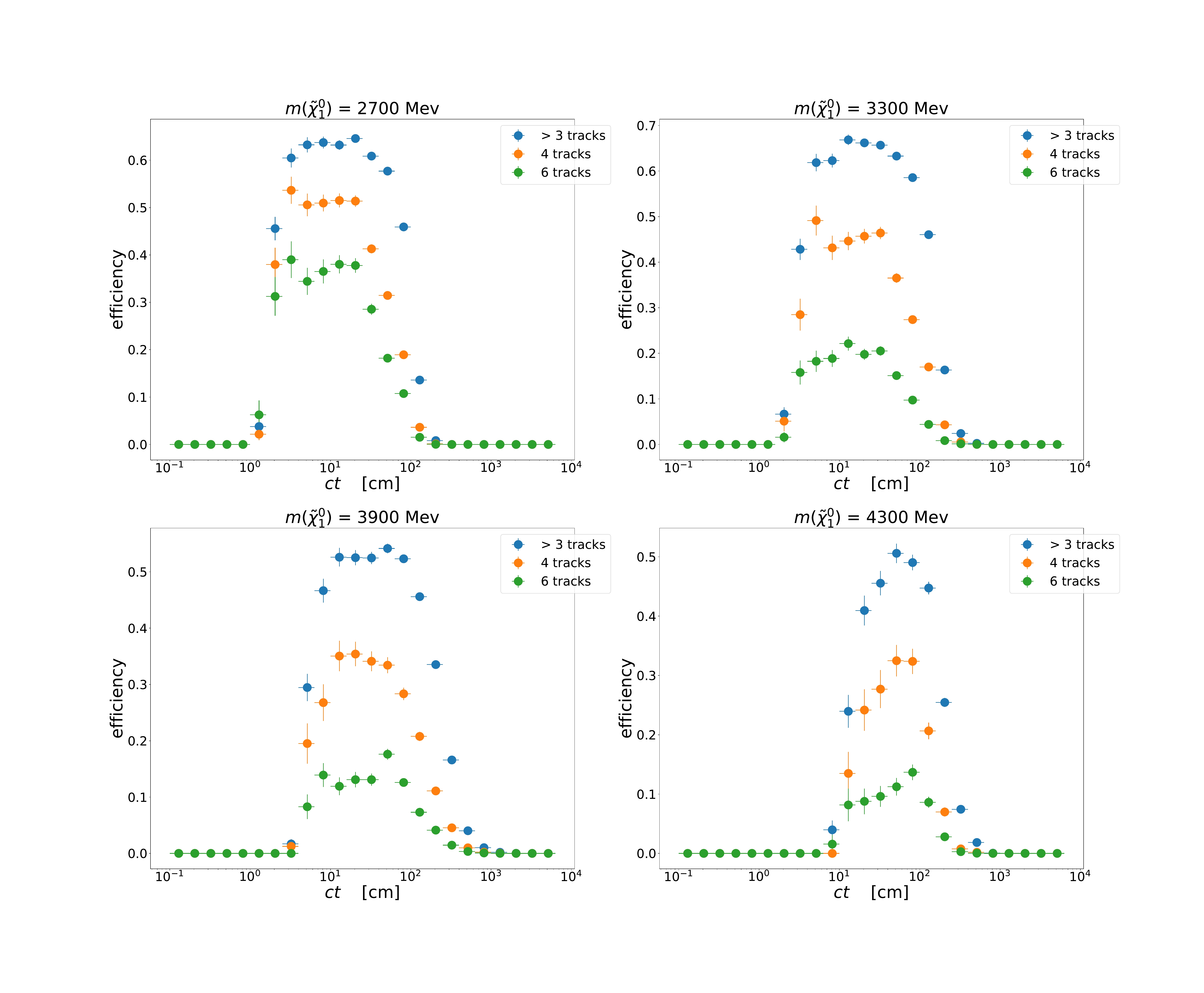}
\caption{Efficiency $\epsilon^{\rm trk}(ct)$ for reconstructing the proton plus the tracks produced in the neutralino decay as a function of the proper decay length $ct$ of the neutralino, shown for selected neutralino masses.
For the inclusive analysis, at least three tracks must be reconstructed, and their invariant mass must be greater than 1.5~GeV (blue).
Lower efficiencies are found for the exclusive-search approaches, where all 4 (orange) or all 6 (green) tracks produced must be detected.
}
\label{fig:eff-vs-ct}    
\end{figure}

The resulting tracking efficiency $\epsilon^{\rm trk}(ct)$ as a function of $ct$, where $t$ is the neutralino decay time in the neutralino rest frame, is shown in Fig.~\ref{fig:eff-vs-ct}, for selected $m_\neu$ values.
We observe that the efficiency is fairly flat over a $ct$ range corresponding to the fiducial volume defined by $r_\DV>1$~cm and the 20-CDC-hit requirement. 
As $m_\neu$ increases, the neutralino boost is smaller, so that the fiducial volume corresponds to larger values of $ct$.

\begin{figure}[t]
    \centering
    \includegraphics[width=0.8\textwidth]{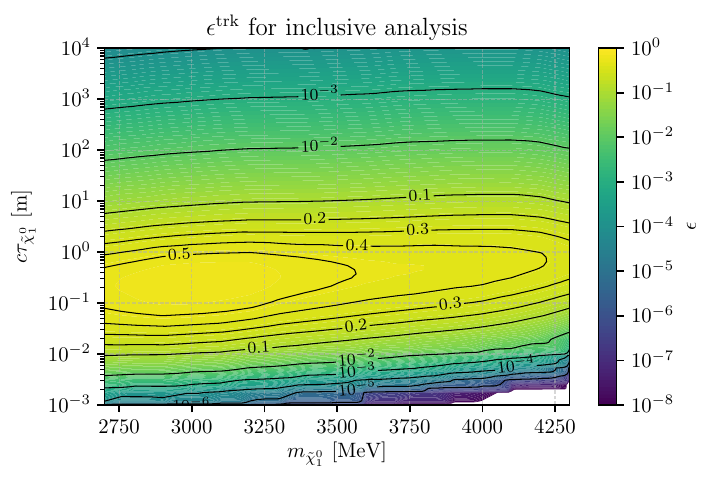}
    \caption{Tracking-reconstruction efficiency at Belle~II, shown in the $(m_{\neu}, \ctau)$ plane, for the inclusive analysis where the lightest neutralino is detected with at least 3 tracks with invariant mass $m_\DV> 1.5$~GeV.
    In the lower-right corner, the efficiency is too small to be determined given the finite size of the simulated sample.} 
    \label{fig:efficiency_inclusive}
\end{figure}

\begin{figure}[t]
    \centering
    \includegraphics[width=0.495\textwidth]{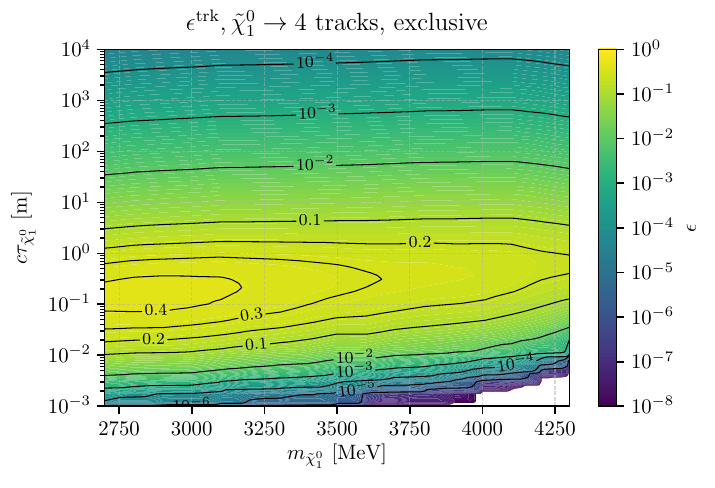}
    \includegraphics[width=0.495\textwidth]{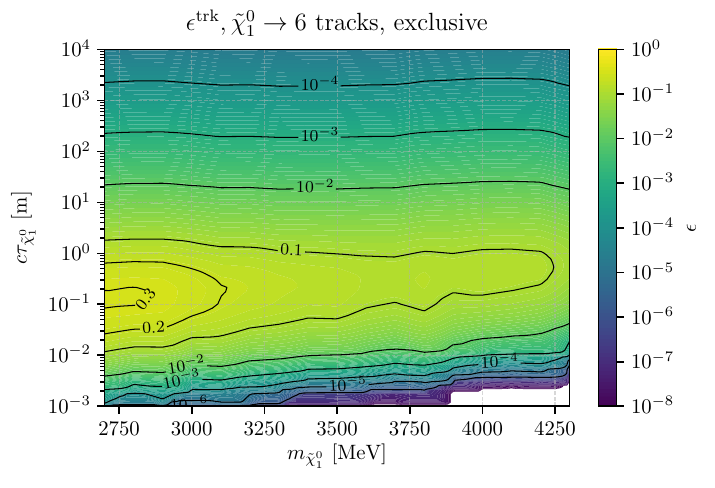}
    \caption{The same as Fig.~\ref{fig:efficiency_inclusive}, but for the exclusive searches with a 4-track (left) and 6-track (right) neutralino-decay final state.}
    \label{fig:efficiency_exclusive}
\end{figure}

Next, we use $\epsilon^{\rm trk}(ct)$ to calculate the tracking efficiency for any value of the average proper decay length $\ctau$,
\begin{eqnarray}
    \epsilon^{\text{trk}}(c\tau_\neu)=\frac{1}{\ctau} \int_0^\infty  d(ct)\, e^{-ct/\ctau}\cdot \epsilon^{\rm trk}(ct).  \label{eqn:efficiency_integration}
\end{eqnarray}
This approach allows for optimal exploitation of the statistics of our simulation sample.
The result of this calculation is presented in Fig.~\ref{fig:efficiency_inclusive} and Fig.~\ref{fig:efficiency_exclusive} for the inclusive and exclusive analyses, respectively, in the $\ctau$ vs.~$m_\neu$ plane.
We observe that generally, the highest efficiency is attained at $\ctau \sim \mathcal{O}(10)$~cm, with this value slightly increasing with mass, largely reflecting the behavior seen in Fig.~\ref{fig:eff-vs-ct}.

\section{Numerical results}
\label{sec:results}

We proceed to calculate the signal-event yields for the inclusive, 4-track exclusive, and 6-track exclusive search analyses, with the following formulas, respectively:
\begin{eqnarray}
    N_S^\text{inclu.}                  &=&2\cdot N_{B^+ B^-}\cdot \text{B}(B^+\to p \neu)\cdot \text{B}(\neu\to \ge 4~\text{tracks})\cdot \epsilon^{\text{trk}} \epsilon^{\text{sol}}, \label{eq:Ninclusive}\\
    N_S^\text{exclu.\&4\text{ trk.}}   &=&2\cdot N_{B^+ B^-}\cdot \text{B}(B^+\to p \neu)\cdot \epsilon^{\text{trk}} \cdot \text{B}(\neu \to \Lambda_c^+ K^- + \text{c.c.})\nonumber \\
    &&\cdot \text{B}(\Lambda_c^+ \to p K^- \pi^+),\\
    N_S^\text{exclu.\&6\text{ trk.}}   &=&2\cdot N_{B^+ B^-}\cdot \text{B}(B^+\to p \neu)\cdot \epsilon^{\text{trk}}\nonumber \\
    &&\cdot \Big(\text{B}(\neu \to \Sigma_c^0 \bar{K}^0 + \text{c.c.}) \cdot  \text{B}(\Sigma_c^0  \to  (\Lambda_c^+\to p K^- \pi^+) \pi^-) \cdot \text{B}(\bar{K}^0 \to \bar{K}_S^0 \to \pi^+ \pi^-)\nonumber\\
    &&+\text{B}(\neu \to \Xi_c^+\pi^-+\text{c.c.})\cdot\text{B}(\Xi_c^+\to \big(\Xi^-\to (\Lambda\to p \pi^-)\pi^-\big)\pi^+\pi^+)\Big),
\end{eqnarray}
where the factors 2 account for the two charged $B$-mesons in each signal process; $N_{B^+B^-}=2.8\times 10^{10}$~\cite{Belle-II:2010dht,Belle-II:2018jsg} is the number of $e^+e^-\to B^+B^-$ events expected with the full Belle~II sample; ``$\text{B}$'' indicates a branching ratio; $\epsilon^{\rm trk}$ is the track-reconstruction efficiency shown in Figs.~\ref{fig:efficiency_inclusive} and \ref{fig:efficiency_exclusive}; and $\epsilon^{\text{sol}}\approx 0.75$ is the efficiency for the cut $D\ge 0$, which ensures that the quadratic equation should have a solution, and for the cut around $m_\pm$ that selects 90\% of the signal of a particular mass.
In Eq.~(\ref{eq:Ninclusive}), $\text{B}(\neu\to \ge 4~\text{tracks})$ is the branching fraction for decay of the neutralino into a final state containing at least 4 tracks and any number of neutral particles, which must take place in order for the decay to satisfy both the requirements of electric-charge conservation and at least 3 detected DV tracks. 
From the simulation, we find that $\text{B}(\neu\to \ge 4~\text{tracks})$ is $0.82$ for $m_\neu=2.7$~GeV and rises to $0.91$ for $m_\neu = 3.4$~GeV.

\begin{figure}[t]
    \centering
    \includegraphics[width=0.8\textwidth]{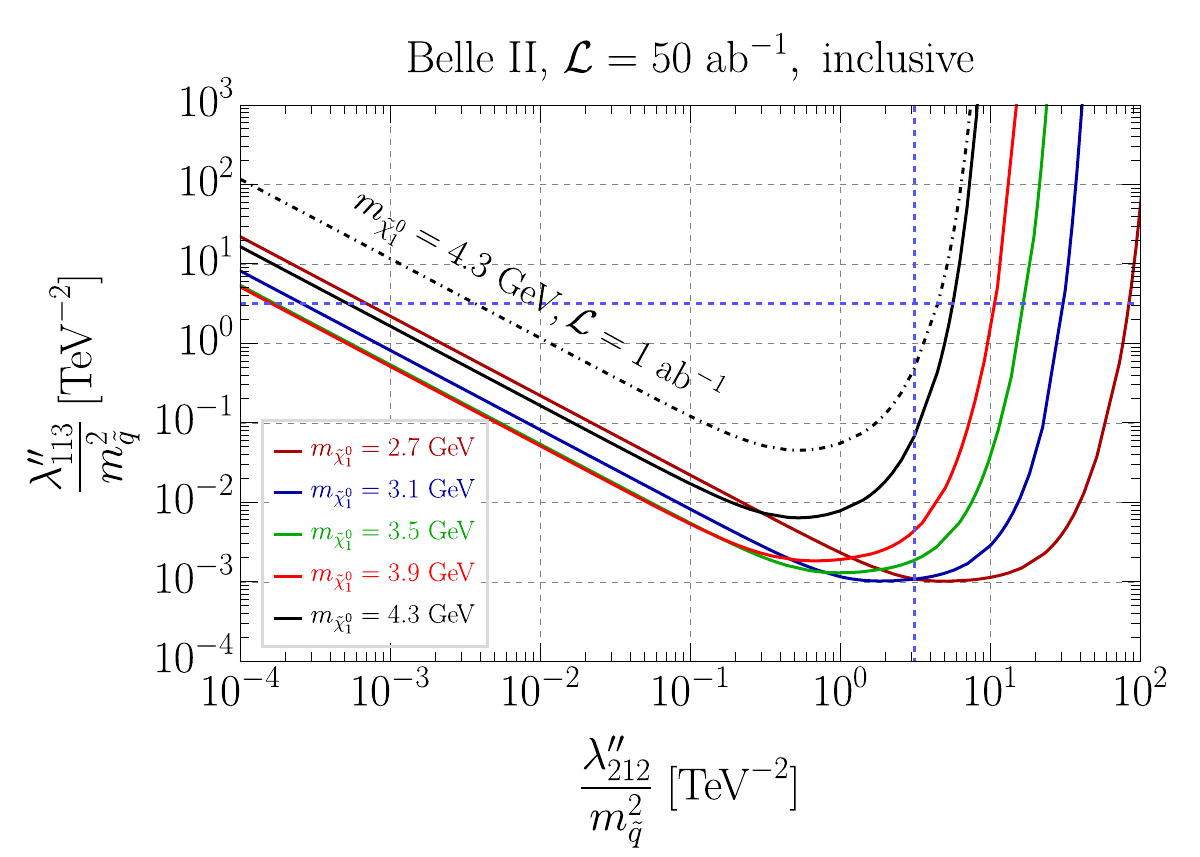}
    \caption{The final sensitivity reach of Belle~II with an integrated luminosity of 50~ab$^{-1}$ and the inclusive analysis, to the light-neutralino scenario, shown in the $\lambda''_{113}/m^2_{\tilde{q}}$ vs.~$\lambda''_{212}/m^2_{\tilde{q}}$ plane for selected values of $m_\neu$ ranging from 2.7 GeV to 4.3 GeV in intervals of 0.4 GeV.
    In addition, we display a dot-dashed curve for $m_{\neu}=4.3$ GeV (black) with an integrated luminosity of 1 ab$^{-1}$.
    The light-blue dotted lines correspond to $\lambda''_{113/212}=4\pi$ for $m_{\tilde{q}}=2$ TeV; for larger values of $\lambda''$ the sensitivity calculation becomes less reliable owing to loss of perturbative unitarity.
    }
    \label{fig:results_inclusive}
\end{figure}

\begin{figure}[t]
    \centering 
    \includegraphics[width=0.495\textwidth]{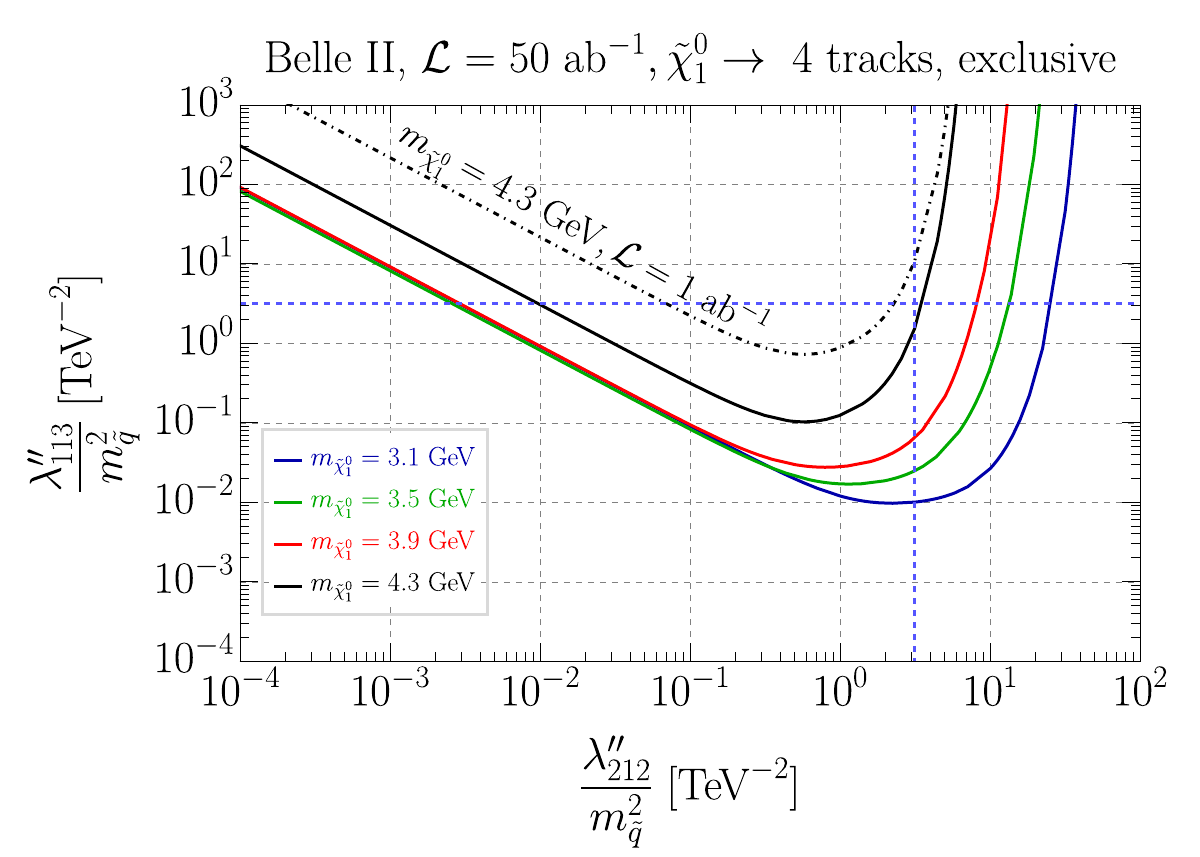}
    \includegraphics[width=0.495\textwidth]{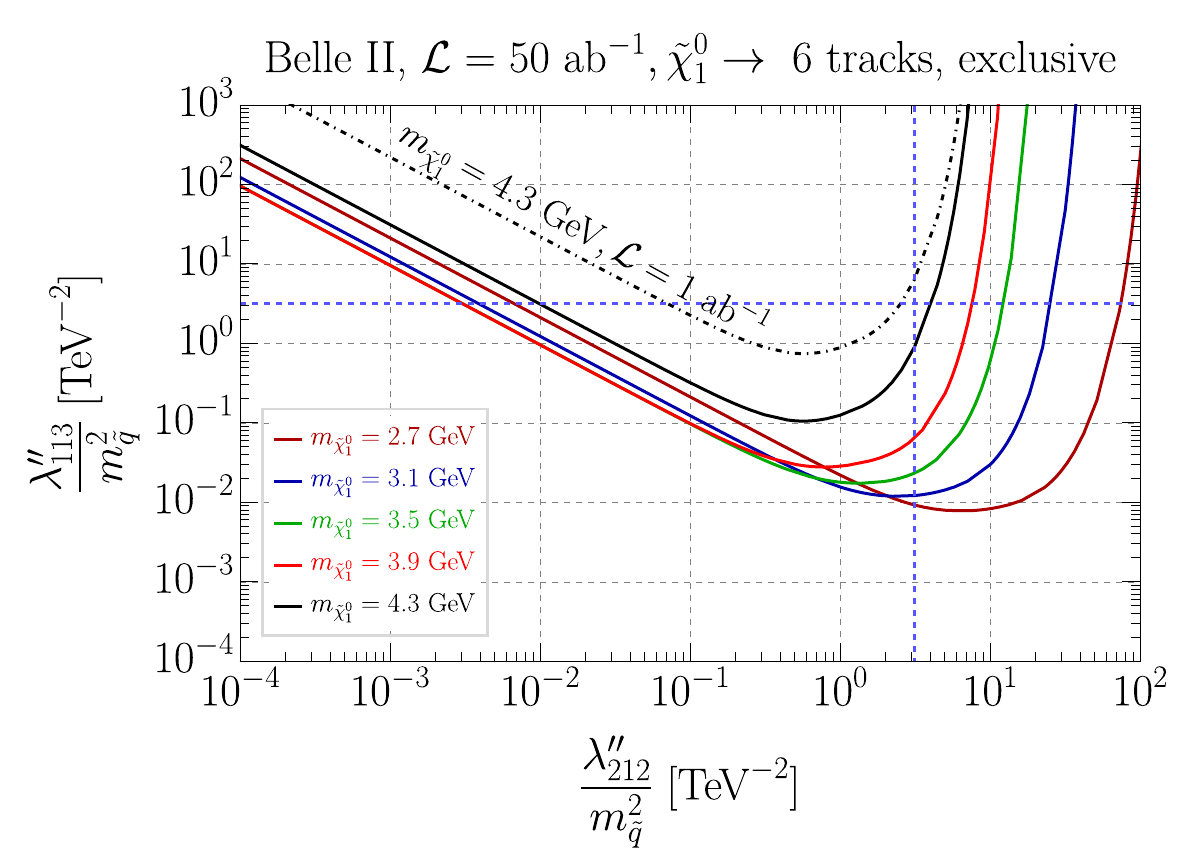}
    \caption{The same as Fig.~\ref{fig:results_inclusive}, but for the two exclusive search analyses.
    The 4-track (left) plot has no curve for $m_\neu=2.7$ GeV, since this mass is kinematically forbidden for the decay $\neu\to \Lambda_c^+ K^-$.}
    \label{fig:results_exclusive}
\end{figure}

As discussed in Sec.~\ref{sec:techniqueANDbackground}, we expect little background with the proposed searches.
Therefore, we take the 95\% confidence-level exclusion limits to lie at parameter values for which we expect to observe 3 signal events.
The corresponding sensitivity reach of Belle~II with an integrated luminosity of 50 ab$^{-1}$ is shown for representative $m_{\neu}$ values in the $\lambda''_{113}/m^2_{\tilde{q}}$ vs.~$\lambda''_{212}/m^2_{\tilde{q}}$ plane in Fig.~\ref{fig:results_inclusive} and Fig.~\ref{fig:results_exclusive} for the inclusive and exclusive analyses, respectively.
In addition, dot-dashed curves show the sensitivity in the example case of $m_{\neu}=4.3$~GeV for an integrated luminosity of $1~ {\rm ab}^{-1}$, which approximately corresponds to the currently available data of Belle and Belle~II.
Comparison of the curves for $1~ {\rm ab}^{-1}$ and $50~ {\rm ab}^{-1}$ shows that, as expected, the sensitivity to $\lambda''_{113}/m^2_{\tilde{q}}$ is proportional to the square root of the integrated luminosity.

These results demonstrate the advantage of the inclusive search, which is sensitive to $\lambda''_{113}/m^2_{\tilde{q}}$ values about a factor of 10 smaller than the exclusive searches.
Furthermore, with $1~{\rm ab}^{-1}$, the inclusive search vastly outperforms the current limits, summarized in Sec.~\ref{subsec:current-limits}.
This includes the best limit on $\lambda''_{113}/m^2_{\tilde{q}}$, set by BABAR with the missing-energy technique~\cite{BaBar:2023dtq}, as long as $\lambda''_{212}/m^2_{\tilde{q}}$ is not smaller than about $10^{-2}~{\rm TeV}^{-2}$.
In addition, taking the BABAR limit into account, we see that the $1~{\rm ab}^{-1}$ inclusive search sensitivity to $\lambda''_{212}/m^2_{\tilde{q}}$ is 1--2 orders of magnitude (depending on $m_{\neu}$) better than the limits obtained from baryon-antibaryon oscillation and dinucleon decays (see Table~\ref{tab:existing_bounds}).
Following the discussion on perturbative unitarity in Sec.~\ref{subsec:current-limits}, we note that sensitivity reach to $\lambda''/m^2_{\tilde{q}}$ greater than about $\mathcal{O}(1~\textrm{TeV}^{-2})$ may not be reliable, given the squark-mass limits.
We thus show light-blue dotted lines in Fig.~\ref{fig:results_inclusive} and Fig.~\ref{fig:results_exclusive} corresponding to $\lambda''_{113/212}=4\pi$ for $m_{\tilde{q}}=2$ TeV.
We should stress that these are not strict bounds but should rather be interpreted as an order-of-magnitude estimate; they indicate that for larger values of $\lambda''$ the sensitivity computation becomes less reliable considering loss of perturbative unitarity.
Nevertheless, we find that Belle II, with an integrated luminosity of either 1 ab$^{-1}$ or 50 ab$^{-1}$, can probe $\lambda''$ couplings up to $\sim 1$--$4$ orders of magnitude below these values.

\section{Conclusions}\label{sec:conclusions}

GeV-scale bino-like neutralinos in RPV SUSY are allowed by all constraints and can be produced in $B$-meson decays in association with a baryon, if BNV operators are switched on.
Further, with such operators, the lightest neutralino $\neu$ can decay to a meson and a baryon.
For a range of values of the coupling and neutralino-mass $m_{\neu}$, the neutralino can be long-lived, such that its decay gives rise to displaced-vertex signatures at collider experiments such as Belle~II.

In this work, we study a representative benchmark scenario that contains two non-vanishing $\lambda''$ couplings, $\lambda''_{113}$ and $\lambda''_{212}$, which multiply $\bar U \bar D \bar D$ operators mediating the production and decay of $\neu$, respectively.
Using the theoretical framework of a phenomenological Lagrangian and the simplified scenario of degenerate squark masses and vanishing squark mixing, we analytically compute the rates for the leading decays of the neutralino, which give rise to a baryon and a meson.
The decay rate of $B^+\to p\neu$ are taken from Ref.~\cite{Dib:2022ppx}.

We propose two strategies to search for a long-lived neutralino in this scenario: inclusive and exclusive.
The inclusive strategy employs a novel partial-reconstruction technique that exploits the displaced vertex produced in the neutralino decay. 
This technique utilizes between 82\% and 91\% of the neutralino decays for $m_\neu=2.7$ and $4.3$~GeV, respectively, resulting in high effective efficiency. 
Furthermore, it enables calculation of the variables $m_\pm$, which correspond to the neutralino mass up to a 2-fold ambiguity. 
We use published searches for other long-lived particles to estimate the background yield from different sources.
We conclude that requiring the displaced vertex to have at least 3 tracks and an invariant mass of $1.5$~GeV is sufficient to bring the background down to the single-event level for any region of $m_\pm$ that corresponds to a particular value of $m_\neu$.
We outline additional background-suppression measures that can be used if needed, at the cost of a reduced efficiency.
The exclusive strategy suppresses background even more effectively, using standard kinematic variables in addition to requiring the presence of the displaced vertex.
Its disadvantage is that it is practical only for specific neutralino decay chains with low total branching fractions, resulting in low effective efficiency.

For both strategies, we perform Monte-Carlo simulations in which we derive detailed track-reconstruction efficiencies using a parameterized model of the Belle~II detector.
Combining these factors, we compute the expected sensitivity of the Belle~II experiment to this scenario, presenting numerical results for three parameters: $\lambda''_{113}/m^2_{\tilde{q}}, \lambda''_{212}/m^2_{\tilde{q}},$ and $m_\neu$.
We find that even with the currently available data of $1~{\rm ab}^{-1}$, the inclusive-strategy search is more sensitive by 1--2 orders of magnitude than the present bounds, obtained from searches for dinucleon decays, baryon-antibaryon oscillations, and $B^+\to p + \text{missing}$.
We also draw lines of $\lambda''_{113/212}=4\pi$ for squark masses of 2 TeV as an order-of-magnitude estimate of perturbative-unitarity considerations; for larger values of $\lambda''_{113/212}$ the sensitivity calculation becomes less reliable owing to loss of perturbative unitarity.
We observe that even in this case, Belle II is estimated to possibly probe the involved RPV couplings multiple orders of magnitude below these values.
Furthermore, the inclusive strategy is more sensitive by about a factor of 10 than the exclusive strategy.

\begin{acknowledgments}

This work was funded by ANID$-$Millen\-nium Program$-$ICN2019\_044 (Chile), ANID PIA/APOYO AFB230003 (Chile), FONDECYT (Chile) under Grants Nos. 1210131, 1241685, 1230160, 1240066 and 1250776, ANID (Chile) FONDECYT Iniciaci\'on Grant No. 11230879, Israel Science Foundation grant 206/23, United States-Israel Binational Science Fund grant 2020044, Horizon 2020 Marie Sklodowska-Curie RISE  project JENNIFER2 Grant Agreement No. 822070, Anusandhan National Research Foundation (ANRF), India grant RJF/2022/00005.
Z.~S.~Wang was supported by the National Natural Science Foundation of China under Grant No.~12475106 and the Fundamental Research Funds for the Central Universities under Grant No.~JZ2025HGTG0252.
Z.~S.~Wang would like to thank the Particle Theory and Cosmology Group of IBS-CTPU, Daejeon, South Korea, for their hospitality, where part of this work was completed.

\end{acknowledgments}

\appendix 
\section{Transition Form Factors} 
\label{app:formfactors} 

The expressions for the hadronic form factors contain two terms: direct contribution (corresponding to direct transition of the neutralino into the final meson-baryon pair) and pole contribution (arising from intermediate contribution of the baryon state strongly decaying into the final meson-baryon pair).
The calculation technique of the hadronic form factors has been discussed in detail in Ref.~\cite{Dib:2022ppx}.

The expressions for the hadronic form factors contain the coupling constant $\beta_{\mathcal{B}}$ defining the matrix element of three-quark operator $\mathcal{O}^{LL}$ between the corresponding baryon state $\mathcal{B}$ and the vacuum~\cite{Dib:2022ppx}.
Our calculation involves five 
$\beta_\mathcal{B}$ 
couplings (see Table~\ref{tab:baryons} for their numerical values), defined as 
\eq
\la \Sigma_c^0|{\cal O}_{ddc}^{LL}|0\ra &=& 
- \bar u_{\Sigma_c^0}(p',s') \, \beta_{\Sigma_c^0}  \, P_L \,, 
\nonumber\\
\la \Omega_c^0|{\cal O}_{ssc}^{LL}|0\ra &=& 
- \bar u_{\Omega_c^0}(p',s') \, \beta_{\Omega_c^0}  \, P_L \,, 
\nonumber\\
\la \Lambda_c^+|{\cal O}_{cud}^{LL}|0\ra &=& 
- \bar u_{\Lambda_c^+}(p',s') \, \beta_{\Lambda_c^+}  \, P_L \,, 
\nonumber\\
\la \Xi_c^+|{\cal O}_{cus}^{LL}|0\ra &=& 
- \bar u_{\Xi_c^+}(p',s') \, \beta_{\Xi_c^+}  \, P_L 
\,,
\nonumber\\
\la \Xi_c^0|{\cal O}_{cds}^{LL}|0\ra &=& 
- \bar u_{\Xi_c^0}(p',s') \, \beta_{\Xi_c^0}  \, P_L \,. 
\label{eq:3quarkops}
\en 

We list below the expressions for the form factors corresponding to the dominant decay modes of the neutralino into a baryon-meson pair induced by the $\lambda''_{212}$ coupling.

Firstly, the form factors for the process $\neu \to \Sigma^0_c + \bar K^0$ are given below as
\eq\label{W0LL_SigcD}
\hspace*{-1cm}
W_0^{LL; \Sigma_c^0 \bar K^0}(p^2)
&=& - \frac{\beta_{\Sigma_c^0}}{f_{K}} \, 
\biggl[ \dfrac{2}{3} \, g^{\tilde cR} \, D \, 
  \Pi_1(m_{\Xi_{c}^{0}},m_{\Sigma^0_c},p^2)
  \nonumber\\
\hspace*{-1cm}
&+& g^{\tilde sR} \, \left(1 
+ \dfrac{D}{3} \, \Pi_1(m_{\Xi_{c}^{0}},m_{\Sigma^0_c},p^2)
+ F \, \Pi_1(m_{\Xi_{c}^{'0}},m_{\Sigma^0_c},p^2)\right) 
\nonumber\\
\hspace*{-1cm}
&+& g^{\tilde dR}  \, \left(- 1 
+ \dfrac{D}{3} \, \Pi_1(m_{\Xi_{c}^{0}},m_{\Sigma^0_c},p^2)
- F \, \Pi_1(m_{\Xi_{c}^{'0}},m_{\Sigma^0_c},p^2)\right) 
      \biggr] \,, 
\en
\eq\label{W1LL_SigcD}   
\hspace*{-1cm}
W_1^{LL; \Sigma_c^0 \bar K^0}(p^2) &=&
- \frac{\beta_{\Sigma^0_c}}{f_{K}} \,  
\biggl[ \dfrac{2}{3} \, g^{\tilde cR} \, D \, 
  \Pi_2(m_{\Xi_{c}^{0}},m_{\Sigma^0_c},p^2)
  \nonumber\\
\hspace*{-1cm}
&+& g^{\tilde sR} \, \left(\dfrac{D}{3} \,
\Pi_2(m_{\Xi_{c}^{0}},m_{\Sigma^0_c},p^2)
+ F \, \Pi_2(m_{\Xi_{c}^{'0}},m_{\Sigma^0_c},p^2)\right) 
\nonumber\\
\hspace*{-1cm}
&+& g^{\tilde dR}  \, \left(
\dfrac{D}{3} \, \Pi_2(m_{\Xi_{c}^{0}},m_{\Sigma^0_c},p^2)
- F \, \Pi_2(m_{\Xi_{c}^{'0}},m_{\Sigma^0_c},p^2)\right) 
\biggr] \,, 
\en
where 
\eq
\Pi_1(m_1,m_2,p^2) &=& 
\dfrac{m_1 m_2 + p^2}{m_1^2 - p^2} 
\,, \nonumber\\ 
\Pi_2(m_1,m_2,p^2) &=& 
\dfrac{m_{\neu} \, (m_1 + m_2)}{m_1^2 - p^2},
\en
and $D = 0.8$, $F = 0.47$~\cite{Dib:2022ppx}. 

The form factors for the process $\neu \to \Omega^0_{c} + K^0$ are obtained from Eqs.~\eqref{W0LL_SigcD} and~\eqref{W1LL_SigcD} upon substitutions
\eq\label{FF_OmegacK}  
m_{\Sigma^0_c} \to m_{\Omega^0_c}\,, \quad
\beta_{\Sigma^0_c} \to \beta_{\Omega^0_c}\,, \quad
D \to - D\,, \quad
g^{\tilde sR} \to g^{\tilde dR}\,, \quad
g^{\tilde dR} \to g^{\tilde sR}\,. 
\en

The form factors parameterizing the matrix element of the $\neu \to \Lambda^+_c + K^-$ process read
\eq\label{W0LL_LamcK}    
\hspace*{-1.75cm}
W_0^{LL; \Lambda^+_c K^-}(p^2) &=& 
-  \sqrt{\dfrac{3}{2}} \,
\frac{\beta_{\Lambda_c^+}}{f_K} \, 
\biggl[ g^{\tilde cR} \, \left(1 - (2 D - F) \, 
  \Pi_1(m_{\Xi_{c}^{0}},m_{\Lambda_{c}^+},p^2)\right) 
\nonumber\\
\hspace*{-1.75cm}
&+& g^{\tilde sR}  \, \left(\frac{1}{2} 
- \dfrac{2 D - F}{2} \, \Pi_1(m_{\Xi_{c}^{0}},m_{\Lambda^+_c},p^2) 
+ \dfrac{D}{2} \, \Pi_1(m_{\Xi_{c}^{'0}},m_{\Lambda^+_c},p^2)\right)  
\nonumber\\
\hspace*{-1.75cm}
&+& g^{\tilde dR}    \, \left(\frac{1}{2} 
  + \dfrac{2 D - F}{2}  \, \Pi_1(m_{\Xi_{c}^{0}},m_{\Lambda^+_c},p^2)  
  + \dfrac{D}{2} \, \Pi_1(m_{\Xi_{c}^{'0}},m_{\Lambda^+_c},p^2)\right)  
\biggr] \,, 
\en

\eq\label{W1LL_LamcK}
\hspace*{-1.75cm}
W_1^{LL; \Lambda_c^+ K^-}(p^2) &=&
-  \sqrt{\dfrac{3}{2}} \,
\frac{\beta_{\Lambda_c^+}}{f_K} \, 
\biggl[ - g^{\tilde cR} \, (2 D - F) \, 
  \Pi_2(m_{\Xi_{c}^{0}},m_{\Lambda_{c}^+},p^2)
\nonumber\\
\hspace*{-1.75cm}
&+& g^{\tilde sR}  \,
\left(- \dfrac{2 D - F}{2} \, \Pi_2(m_{\Xi_{c}^{0}},m_{\Lambda^+_c},p^2) 
+ \dfrac{D}{2} \, \Pi_2(m_{\Xi_{c}^{'0}},m_{\Lambda^+_c},p^2)\right)  
\nonumber\\
\hspace*{-1.75cm}
&+& g^{\tilde dR}    \, \left( 
    \dfrac{2 D - F}{2}  \, \Pi_2(m_{\Xi_{c}^{0}},m_{\Lambda^+_c},p^2)  
  + \dfrac{D}{2} \, \Pi_2(m_{\Xi_{c}^{'0}},m_{\Lambda^+_c},p^2)\right)  
  \biggr] \,. 
\en

Finally, the form factors for the processes $\neu \to \Xi_{c}^{+} + \pi^-$ and $\neu \to \Xi_{c}^{0} + \pi^0$ 
are derived from Eqs.~\eqref{W0LL_LamcK} and~\eqref{W1LL_LamcK} upon the following substitutions 
\eq\label{FF_Xicaspi1}  
m_{\Lambda_{c}^+} \to m_{\Xi_{c}^{+}}\,, \quad
\beta_{\Lambda_{c}^+} \to \beta_{\Xi_{c}^{+}}\,, \quad
f_K \to f_\pi,
\en
and 
\eq\label{FF_Xicaspi2}  
m_{\Lambda_{c}^+} \to m_{\Xi_{c}^{0}}\,, \quad
\beta_{\Lambda_{c}^+} \to \beta_{\Xi_{c}^{0}}\,, \quad
f_K \to f_\pi\,,
\en
respectively.
In the case of the $\neu \to \Xi_{c}^{0} + \pi^0$ mode, one should include the isospin factor $-1/\sqrt{2}$ in both form factors.

\section{Partial-reconstruction calculation of the neutralino-candidate mass}
\label{app:mass}

Consider a two-body decay such as $B^+\to p\neu$, where one of the daughter particles ($\neu$) is long lived.
The decay is reconstructed partially: in addition to the proton, only two or more charged-particle tracks from the neutralino decay are reconstructed and used to determine the position of the DV, yielding the neutralino flight direction $\hat p_\neu$.
We write the three-momentum conservation in the $B^+$ decay as
\begin{equation}
    \vec p_B = \vec p_p + p_\neu \hat p_\neu .
    \label{eq:p_B}
\end{equation}
Note that this equation holds in the laboratory frame, since the DV position, and hence $\hat p_\neu$, is known only in that frame.
The $z$-component of the $B^+$ momentum is
\begin{equation}
    p_B^z = p_p^z + p_\neu \cos \theta_\neu ,
\label{eq:p_B^z}
\end{equation}
where the $z$ direction is taken to be that of the velocity of the CM frame in the laboratory frame and 
$\theta_\neu$ is the measured polar angle of $\hat p_\neu$.

Similarly, energy conservation in the laboratory frame gives
\begin{equation}
    E_B = E_p + E_\neu .
\label{eq:E_B}
\end{equation}
The energy of the $B^+$ meson in the CM frame, which equals half the CM energy $\sqrt{s}/2$, is related to $E_B$ by the Lorentz transformation
\begin{equation}
    \frac{\sqrt{s}}{2} = \gamma \left(E_B - \beta (p_p^z + p_\neu \cos \theta_\neu)\right),
    \label{eq:sqrts}
\end{equation}
where $\beta$ and $\gamma$ give the Lorentz boost between the two frames, and $p_B^z$ was replaced with Eq.~(\ref{eq:p_B^z}).

Lastly, the known $B^+$ mass can be written in terms of the laboratory-frame quantities
\begin{equation}
    m_B^2 = E_B^2 - \left(p_p^2 + p_\neu^2 + 2 p_p p_\neu \cos \theta_{p\neu}\right), 
\label{eq:mass}
\end{equation}
where $\theta_{p\neu}$ is the measured angle between $\vec p_p$ and $\hat p_\neu$.

Both Eqs.~\eqref{eq:sqrts} and~\eqref{eq:mass} provide a relation between $E_B$ and $p_\neu$, and hence can be put in the form of a single quadratic equation in terms of $p_\neu$.
The quadratic equation has two solutions.
For each solution, use of Eqs.~(\ref{eq:sqrts}), (\ref{eq:E_B}), and (\ref{eq:p_B}) yields the four momenta of both the $\neu$ and the $B^+$. 
In particular, one obtains two solutions for the squared neutralino mass, $m_\neu^2 = E_\neu^2 - p_\neu^2$.

\bibliographystyle{JHEP}
\bibliography{main}

\end{document}